\documentclass[10pt,superscriptaddress,twocolumn,aps,showpacs,nofootinbib,eqsecnum]{revtex4}

\usepackage[pdfborder={0 0 0},colorlinks=true,linkcolor=blue,citecolor=blue %
,filecolor=blue,urlcolor=blue]{hyperref}   
\usepackage{url}
\usepackage{ulem}
\usepackage{slashed}
\usepackage{amssymb}
\usepackage{amsfonts}
\usepackage{mathbbol} 
\usepackage{amstext}
\usepackage{graphicx}
\usepackage{color}
\usepackage{epsfig,amsmath,lscape}

\newcommand{\myindent}{\hspace{15pt}}

\begin{document}

\title{Neutron star properties and the symmetry energy}

\author{Rafael Cavagnoli} 
\affiliation{Departamento de F\'{\i}sica - CFM - Universidade Federal de Santa Catarina, \\ 
Florian\'opolis - SC - CP. 476 - CEP 88.040 - 900 - Brazil}
\affiliation{Centro de F\'{i}sica Te\'{o}rica, Departamento de F\'{i}sica,
Universidade de Coimbra, P-3004-516 Coimbra, Portugal}

\author{Constan\c{c}a Provid\^encia}
\affiliation{Centro de F\'{i}sica Te\'{o}rica, Departamento de F\'{i}sica,
Universidade de Coimbra, P-3004-516 Coimbra, Portugal}

\author{Debora P. Menezes}
\affiliation{Departamento de F\'{\i}sica - CFM - Universidade Federal de Santa Catarina, \\ 
Florian\'opolis - SC - CP. 476 - CEP 88.040 - 900 - Brazil}

\begin{abstract}
The effect of the symmetry energy on the properties of compact stars is
discussed. It is shown that, for stars with masses above 1 $M_\odot$, the radius
of the star varies linearly with the symmetry energy slope $L$. The
dependence of the hyperon content and onset density of the direct Urca process on the symmetry energy and meson coupling
parametrization are also analyzed. 
\end{abstract}


\maketitle

\section{Introduction}

\myindent 
Constraining the high density equation of state (EOS) of neutron rich matter is
essential to understand the physics of compact stars \cite{lattimer07}.

\myindent There have been recent attempts to set constraints on the high density EOS
using  observational data obtained from compact stars \cite{ozel09,steiner10,hebeler10}. 
In particular, in \cite{steiner10} an empirical dense equation of
state obtained  from  a heterogeneous set of six
neutron stars with well-determined distances was proposed.

\myindent Phenomenological nuclear models are generally fitted to the ground-state
properties of nuclei and, frequently, also to the collective response of these
systems \cite{nl3,fsu,iufsu}, or nuclear matter saturation properties
\cite{gm1}. However, these constraints generally
only determine quite uniquely the EOS close to saturation density and for an isospin
asymmetry smaller than 0.2 \cite{rafael11,camille11}. Extrapolation to high densities and/or high
isospin asymmetries is kept unconstrained and different models predict quite
different neutron star properties.

\myindent In \cite{pieka2002} it was proposed that the parametrization of the
nuclear EOS could also be constrained by the collective response of nuclei to the
isoscalar monopole giant resonance (ISGMR)  and the isovector dipole giant
resonance (IVGDR). The author of  \cite{pieka2002} has proposed that the ISGMR and IVGDR of
$^{208}$Pb were sensitive  both to the incompressibility $K$ and the symmetry
energy $\epsilon_{sym}$, due to its isospin asymmetry. Therefore,  the ISMGR
data  from a  nucleus with a  well
developed breathing mode but a small  neutron-proton asymmetry  such as
$^{90}$Zr should be used to fix  the incompressibility at
saturation instead of a nucleus with a non negligible isospin asymmetry like
$^{208}$Pb. Once the incompressibility at saturation is fixed, the   IVGDR
$^{208}$Pb may be used to constrain the symmetry energy. 

\myindent This information together with the ground-state properties of nuclei has been
used to define the FSU parametrization proposed in \cite{fsu}. However, since
the high density EOS is not constrained, FSU presents an EOS
that is too soft at high densities and does not predict a star with a mass
larger than 1.72 $M_\odot$, $0.25\, M_\odot$ below the mass 1.97$\pm0.04\, M_{\odot}$  of the recent mass measurement 
of the binary millisecond pulsar PSR J1614−2230 \cite{demorest10}. In order to
overcome this drawback, the parametrization \cite{iufsu} was built in a way
close to FSU but including an
extra constraint: the EOS is compatible with the empirical equation of
state determined in \cite{steiner10}. As a result the new parametrization 
predicts stars with larger masses and smaller radii \cite{iufsu}.

\myindent In the present work we want to understand how sensitive is the mass/radius
curve of a family of stars to the symmetry energy and its slope at
saturation. We study not only maximum mass configurations but also stars
with a mass in the range $1.0\, M_\odot< M< 1.4\, M_\odot$. These stars  have a
central density that goes from 1.5 $\rho_0$ to 2-3 $\rho_0$, and therefore we
will be testing the equation of state at suprasaturation densities. 

\myindent At high density the formation of hyperons is energetically favorable and
therefore we also study the effect of the symmetry energy on the
appearance of this exotic degrees of freedom \cite{gm1,glen00}. We 
consider two different hyperon-meson parametrizations: one proposed in \cite{gm1}
and a second one that takes into account
the different binding energies of the hyperons \cite{hyp1,chiap09}.

\myindent In section \ref{formalism} we present the formalism used in the present
work, in section \ref{results} the results are presented and discussed and
in the last section conclusions are drawn.


\section{The Formalism}
\label{formalism}

\myindent  
%

\myindent In the present section we present the hadronic equations of state (EOS) 
used in this work. We describe hadronic matter
within the framework of the relativistic non-linear Walecka model (NLWM)
\cite{bb}. In this  model the nucleons are coupled to 
neutral scalar $\sigma$, 
isoscalar-vector $\omega_\mu$ and isovector-vector $\vec \rho_\mu$  meson fields. 
We include a $\rho-\omega$ meson coupling term as in \cite{hor01,fsu,iufsu} in
order to study the effect of the symmetry energy on the star
properties while leaving the isoscalar channel fixed.

\myindent The Lagrangian density reads

\begin{eqnarray} 
{\cal L}&=& \sum\limits_{j{\kern 1pt}  = {\kern 1pt} 1}^8 {\bar \psi_j \left[ ~ { \gamma_\mu  \left( {i \partial^\mu 
- g_{\omega j} \, \omega^\mu - g_{\rho j} \, \vec \tau _j \,.\, \vec \rho^{\, \mu}  } \right) 
- m_j^* ~} \right]\psi_j }  \nonumber \\ 
&+& \sum\limits_{l{\kern 1pt}  = {\kern 1pt} 1}^2 {\bar \psi _l \left( {i\gamma _\mu  \partial ^\mu 
- M_l } \right)\psi _l}  \nonumber \\ 
&+& \frac{1}{2} {\partial_\mu \sigma \partial^\mu \sigma - \frac{1}{2} m_{\sigma}^2 \sigma^2 }  
- \frac{1}{{3!}} k \sigma^3  - \frac{1}{{4!}} \lambda \sigma^4     \nonumber \\
&-& \frac{1}{4} \Omega_{\mu \nu} \, \Omega^{\mu \nu}
+ \frac{1}{2} m_{\omega}^2 \, \omega_\mu \omega^\mu 
+ \frac{1}{4!}\xi g_{\omega}^4 (\omega_{\mu} \omega^{\mu})^2  \nonumber \\
&-& \frac{1}{4} \vec R_{\mu \nu } \,.\,  \vec R^{\mu \nu } + 
\frac{1}{2} m_\rho^2 \, \vec \rho_{\mu} \,.\, \vec \rho^{\, \mu}   \nonumber \\
&+& \Lambda_{\rm v} (g_{\rho}^2 \; \vec \rho_{\mu} \,.\, \vec \rho^{\, \mu} )(g_{\omega}^2 \; \omega_{\mu} \omega^{\mu} )  \;,   
%
\label{baryon-lag}   
\end{eqnarray}
%
$ $

where $m_j^* = m_j - g_{\sigma j}\, \sigma$ is the baryon effective mass,  
$\Omega_{\mu\nu }=\partial_\mu \omega_\nu - \partial_\nu \omega_\mu$~, 
$\vec R_{\mu \nu } = \partial_\mu \vec \rho_\nu -\partial_\nu \vec \rho_\mu 
-g_\rho \left({\vec \rho_\mu \,\times \, \vec \rho_\nu } \right)$, $g_{ij}$ 
are the coupling constants of mesons $i = \sigma, \omega, \rho$ with baryon 
$j$, $m_i$ is the mass of meson $i$ and $l$ represents the leptons $e^-$ and $\mu^-$. 
The couplings 
$k$~($k = 2\,M_N\,g_{\sigma}^3\,b$) and $\lambda$~($\lambda = 6\, g_{\sigma}^4\,c$) 
are the weights of the non-linear scalar terms and $\vec \tau$ is the isospin
operator.
%
%
The sum over $j$ in (\ref{baryon-lag}) extends over the octet of lightest baryons 
$\{ n,p,\Lambda,\Sigma^-,\Sigma^0,\Sigma^+,\Xi^-,\Xi^0 \}$. 

\myindent We consider two different sets of hyperon-meson couplings. For the coupling
set A the $\omega$ and $\rho$ meson-hyperon coupling constants are obtained 
using the SU(6) symmetry:
%
%
\begin{equation}
 \frac{1}{2} g_{\omega \Lambda} = \frac{1}{2} g_{\omega \Sigma} = g_{\omega \Xi} = \frac{1}{3} g_{\omega N}   \:,
\label{omega-h}
\end{equation}
\begin{equation}
\frac{1}{2} g_{\rho \Sigma} = g_{\rho \Xi} = g_{\rho N}  \quad ; \quad  g_{\rho \Lambda} = 0    \:,
\label{rho-h}
\end{equation}

where $N$ means 'nucleon' $(g_{iN} \equiv g_{i})$. The coupling constants 
$\{ g_{\sigma j} \}_{j=\Lambda,\Sigma,\Xi}$ of the hyperons with the scalar 
meson $\sigma$ are constrained
by the hypernuclear potentials in nuclear matter to be consistent with hypernuclear 
data \cite{hyp1}. The hypernuclear potentials were constructed as
\begin{equation}
V_j = x_{\omega j} \, V_{\omega} - x_{\sigma j} \, V_{\sigma}    \:,
\label{hyppot}
\end{equation}
where $x_{ij} \equiv g_{ij}/g_{i}$, $V_{\omega} \equiv g_{\omega} \omega_0$~and 
$V_{\sigma} \equiv g_{\sigma} \sigma_0$~are the nuclear potentials for symmetric 
nuclear matter at saturation with the 
parameters of Table \ref{tab-meson-pot}. Following Ref. [33] we use
\begin{equation}
V_{\Lambda} = -28~{\rm MeV} \;\;,\;\; V_{\Sigma} = 30~{\rm MeV} \;\;,\;\;  
V_{\Xi} = -18~{\rm MeV}  \:.
\end{equation}
All hyperon coupling ratios $\{ g_{\sigma j} , g_{\omega j} , g_{\rho j} \}_{j=\Lambda,\Sigma,\Xi}$ 
are now known once the coupling constants $\{g_{\sigma}, g_{\omega}, g_{\rho} \}$ 
of the nucleon sector are given.

\begin{table}[ht]
\centering
\begin{tabular}{ccccccc}
\hline
 & \, NL3 \,\, &  \, GM1 \,  &  \, GM3 \,  &  \, NL$\rho$ \,  &  \; FSU \;  &  IU-FSU  \\ 
\hline
$V_{\sigma}$~(MeV) & 377.15 & 281.34 & 206.28 & 234.68 & 358.94 & 359.15 \\ 
$V_{\omega}$~(MeV) & 305.46 & 215.71 & 145.45 & 171.10 & 282.42 & 276.84 \\ 
\hline
\end{tabular}
\caption{The $\sigma$ and $\omega$ meson potentials for symmetric nuclear matter at saturation.}
\label{tab-meson-pot}
\end{table}

\myindent However, while the binding of the $\Lambda$ to symmetric nuclear matter is
well settled experimentally, the  binding values of the $\Sigma^-$ and $\Xi^-$
have still a lot of uncertainties. Therefore, in order to show how results are
sensitive to the hyperon couplings we consider set B defined as proposed in
\cite{gm1} with  $x_\sigma$=0.8 and equal for all the hyperons. We obtain the
fraction $x_\omega$ from eq. (\ref{hyppot}) with $V_j=V_{\Lambda} = -28~{\rm MeV}$,
and take the same value for all the hyperons. For the hyperon-$\rho$-meson
coupling we consider $x_\rho=x_\sigma$. 
This choice of coefficients has shown to give high maximum mass
configurations \cite{bombaci08} which could describe the
millisecond pulsar J1614-2230  mass \cite{demorest10}.   

\myindent In Table \ref{tab-parameters} we give the symmetric nuclear matter properties 
at saturation density as well as the parameters of the models used in the 
present work.

{
\myindent The equations of state (EOS) are the standard relativistic mean field equations  
known in the literature. In case of eq. (\ref{baryon-lag}), the EOS are the 
same as presented in \cite{rafael11} for the hadronic case.
The baryon number density is:
\begin{equation}
 n_B =  \sum_{j = 1}^{8} { n_j  }  \; ,
\label{nb-hyp}
\end{equation}
where $n_j$ is the baryon number density of baryon $j$ at zero temperature:
\begin{equation}
 n_j =  \frac{1}{3 \pi^2} k_{F j}^3  \; ,
\end{equation}
and $k_{F j}$ is the Fermi momentum of baryon $j$. For the sake of comparisons with 
symmetric nuclear matter ($n_B = n_p + n_n$), the symmetry energy is defined as:
\begin{equation}
 {\cal E}_{sym} = \frac{1}{2} \left [ \frac{\partial^2 ({\cal
       E}/n_B)}{\partial \alpha^2}  \right ]_{\alpha = 0}=
\frac{k_F^2}{6 E_F}+\frac{g_\rho^2}{4{m^*_\rho}^2}n_B
\; ,
\end{equation}
where ${\cal E}$ is the energy density obtained from eq. (\ref{baryon-lag}) for $j = 1,2$ and with no leptons, 
$\alpha$ is the asymmetry parameter $\alpha = (N - Z)/A = (n_n - n_p)/n_B$,
$E_F=(k_F^2+{m^*}^2)^{1/2}$ with 
$k_F=(3\pi^2 n_B/2)^{1/3}$, ${m^*_\rho}^2=m_\rho^2 +2\Lambda_{\rm v} g_\omega^2 g_\rho^2
\omega_0^2$, 
and the slope of the symmetry energy is:
\begin{equation}
 L = \left [ 3 n_B \frac{\partial {\cal E}_{sym}}{\partial n_B}   \right ]_{n_B = n_0} \; .
\end{equation}

When the hyperons are present we define the strangeness fraction:
\begin{equation}
 f_s = \frac{1}{3} \frac{\sum_{j}^{} |s_j| n_j }{n_B} \; .
\end{equation}

where $s_j$ is the strangeness of baryon $j$ and $n_B$ in this case 
is given by eq. (\ref{nb-hyp}). 

}

%
\section{Results}
\label{results}
\begin{table*}[ht]
\centering
\begin{tabular}{ccccccc}
\hline
 &  {\bf FSU} \cite{fsu}  &  {\bf IU-FSU} \cite{iufsu}   &  {\bf NL$\rho$} \cite{nlrho}  &  {\bf NL3} \cite{nl3}  &  {\bf GM1} \cite{gm1}  &  {\bf GM3} \cite{gm1} \\ 
\hline
$n_0$ (fm$^{-3}$)          &    0.148   &   0.155   &   0.160   &   0.148   &   0.153   &  0.153  \\ 
$K$ (MeV)                  &    230     &   231.2   &   240     &   271.76  &   300     &   240   \\ 
$m^*/m$                    &    0.62    &   0.62    &   0.75    &   0.60    &   0.70    &   0.78   \\ 
$m$~(MeV)                  &    939     &   939     &   939     &   939     &   938     &   938    \\ 
-$B/A$ (MeV)               &    16.3    &   16.4    &   16.0    &   16.299  &   16.3    &   16.3   \\ 
${\cal E}_{\rm sym}$ (MeV) &    32.6    &   31.3    &   30.5    &   37.4    &   32.5    &   32.5   \\ 
$L$ (MeV)                  &    61      &   47.2    &   85      &   118     &    94     &   90     \\
\hline
$m_{\sigma}$ (MeV)         &    491.5   &   491.5   &   512     &  508.194  &   512     &   512    \\ 
$m_{\omega}$ (MeV)         &    782.5   &   782.5   &   783     &   783     &   783     &   783    \\ 
$m_{\rho}$ (MeV)           &    763     &   763     &   763     &   763     &   770     &   770    \\ 
$g_{\sigma}$               &    10.592  &   9.971   &   8.340   &  10.217   &  8.910    &  8.175   \\ 
$g_{\omega}$               &    14.302  &   13.032  &   9.238   &  12.868   &  10.610   &  8.712   \\ 
$g_{\rho}$                 &    11.767  &   13.590  &   7.538   &  8.948    &  8.196    &  8.259   \\ 
$b$                        & \;\; 0.000756 \;\;  & \;\; 0.001800 \;\;  & \;\; 0.006935 \;\; &\;\; 0.002052 \;\; & \;\; 0.002947 \;\; & \;\; 0.008659  \\ 
$c$                        &  0.003960  & 0.000049  & -0.004800 & -0.002651 & -0.001070 & -0.002421 \\ 
$\xi$                      &    0.06    &   0.03    &     0     &     0     &     0     &    0     \\ 
$\Lambda_{\rm v}$          &    0.03    &   0.046   &     0     &     0     &     0     &    0     \\ 
\hline
\end{tabular}
\caption{Parameter sets used in this work and corresponding saturation 
properties.}
\label{tab-parameters}
\end{table*}

\begin{figure}[ht]
  \centering
\begin{tabular}{cc}
\includegraphics[width=8cm,height=6.0cm,angle=0]{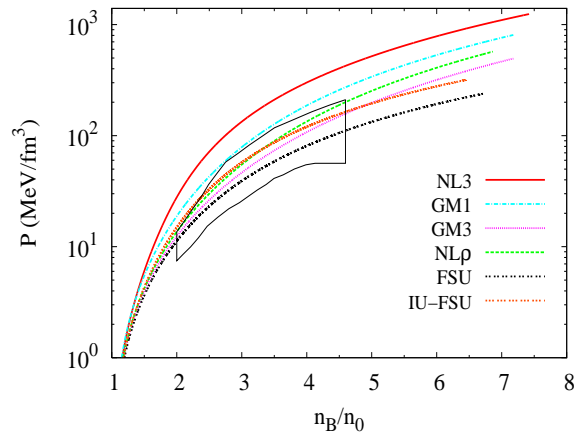}
\end{tabular}
\caption{(Color online) EOS for symmetric matter and different models: pressure as a function of the baryon number 
density. The enclosed area represents experimental data according to Danielewicz {\it et al.}, \cite{danielewicz1}. }
 \label{press}
\end{figure}

\begin{figure*}[ht]
  \centering
\begin{tabular}{cc}
\includegraphics[width=8cm,height=6.0cm,angle=0]{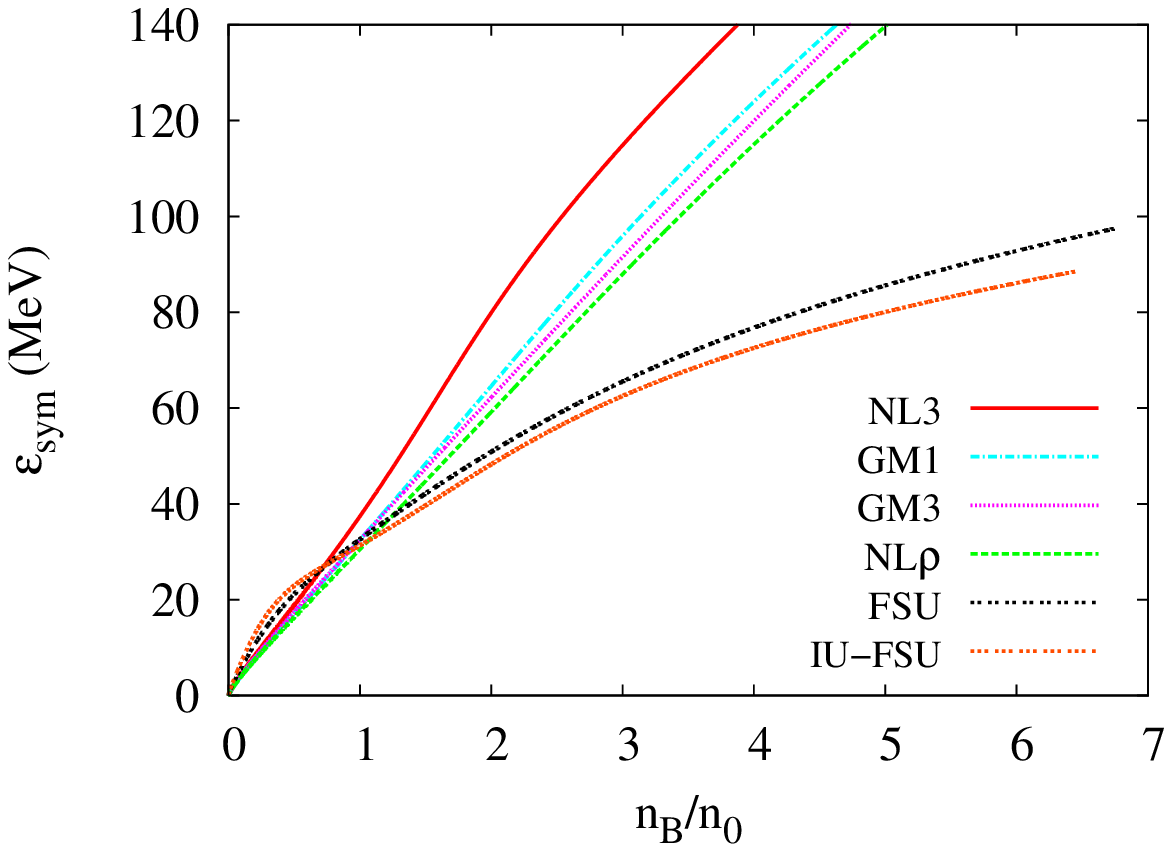}  
 &
\includegraphics[width=8cm,height=6.0cm,angle=0]{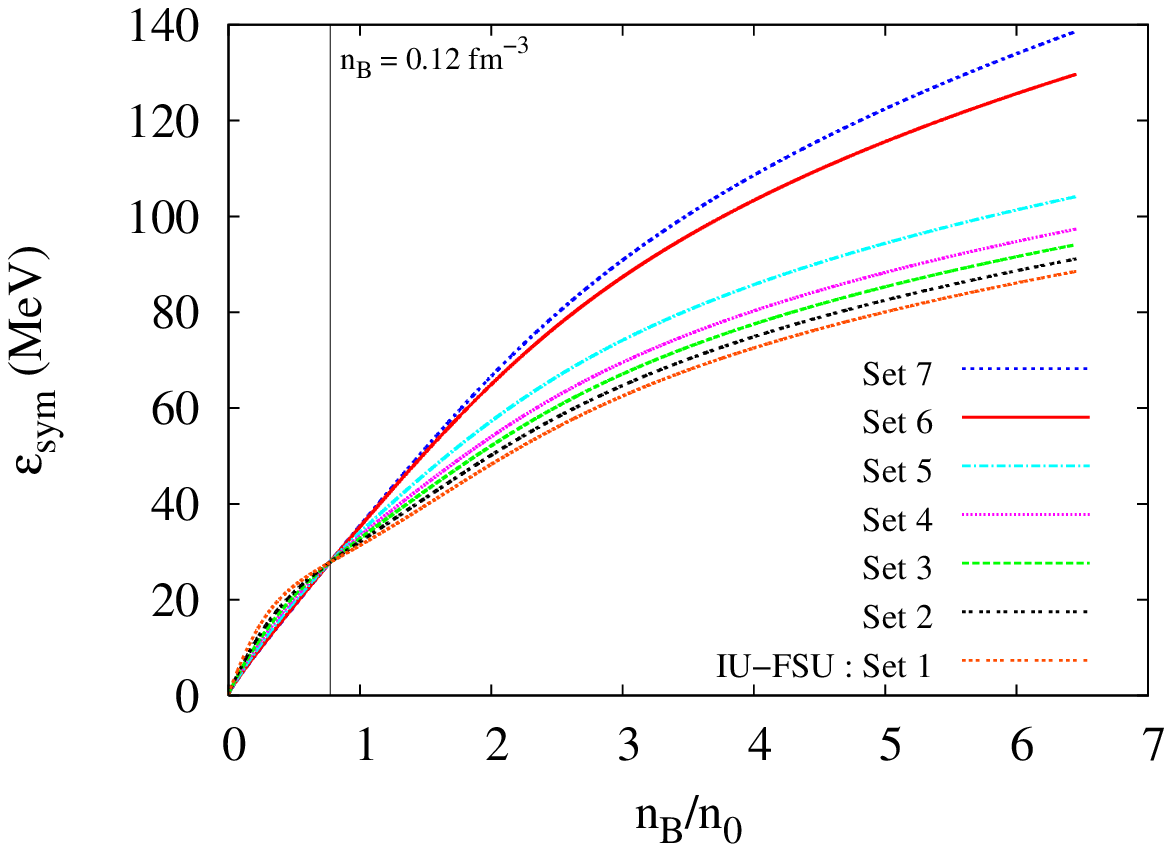}  
\\ {\bf (a)} & {\bf (b)}
\end{tabular}
\caption{(Color online) The symmetry energy as a function of the baryon number density a)
for  different models; b) for modified IUFSU.  }
 \label{ener-sym}
\end{figure*}

In Figs.  ~\ref{press} and \ref{ener-sym}a)  the pressure of symmetric nuclear matter and
the symmetry energy, respectively, are  plotted for a large range of
densities. In  Fig.~\ref{press}
  we also include the experimental constraints   obtained from
collective flow data in
heavy-ion collisions  \cite{danielewicz1}. We have considered a wide range of models frequently
used to study stellar matter or finite nuclei with quite different behaviours
at high densities. Even though some of the models
do not satisfy the  constraints
determined in  \cite{danielewicz1}, as a whole this set of models allows us 
to understand
the influence of a hard/soft equation of state (EOS) and a hard/soft symmetry
energy on the star properties.  

\myindent We have considered the following parametrizations:
NL3 \cite{nl3}, with a quite large symmetry energy and
incompressibility at saturation and which was fitted in order to
reproduce the ground state properties of both stable and unstable
nuclei. FSU \cite{fsu}, which was
accurately calibrated to simultaneously describe the GMR in
$^{90}$Zr and $^{208}$Pb, and the IVGDR in $^{208}$Pb and still
reproduce ground-state observables of stable and unstable nuclei. FSU is very
soft at high densities, therefore the authors of \cite{iufsu} have proposed a
parametrization with similar properties, which they call IUFSU, having a harder behavior at high densities.
GM1 and GM3 \cite{gm1} are generally used to describe stellar matter, with a 
symmetry energy not so hard as the one of NL3, and 
NL$\rho$ \cite{nlrho}, which has been used to discuss the hadron matter-quark matter
transition in \cite{ditoro2010}, and that has, at high densities, a behavior between GM1 and GM3.

\begin{table*}[ht]
\centering
\begin{tabular}{cccccccc}
\hline
 & \;\;\; Set 1 \;\;\; & \;\;\; Set 2 \;\;\; &  \;\;\; Set 3 \;\;\;  &  \;\;\; Set 4 \;\;\;  &  \;\;\; Set 5 \;\;\;  &  \;\;\; Set 6 \;\;\;  &  \;\;\; Set 7 \;\;\;   \\ 
\hline
$g_{\rho}$ & 13.590 & 11.750 & 10.750 & 10.150 & 9.500 & 8.750 & 8.650 \\ 
$\Lambda_{\rm v}$ & 0.0406 & 0.03643 & 0.02905 & 0.02354 & 0.01635 & 0.00598 & 0.00439 \\ 
\hline
${\cal E}_{\rm sym}$~(MeV)  & 31.34 & 32.09 & 32.74 & 33.26 & 34.00 & 35.21 & 35.41 \\ 
$L$~(MeV) & 47.20 & 55.09 & 62.38 & 68.73 & 78.45 & 96.02 & 99.17 \\ 
\hline
\end{tabular}
\caption{Parameter sets generated from the IU-FSU model (Set 1) that differ in their value of the symmetry energy 
${\cal E}_{\rm sym}$ and corresponding slope $L$ at saturation but have the same isoscalar properties.}
\label{tab-sets-grho}
\end{table*}

\myindent In order to study the effect of the isovector channel in the star properties
we also consider a modified version of the IUFSU parametrization: we keep the
isoscalar channel and change $g_\rho$ and $\Lambda_{\rm v}$ keeping the symmetry
energy fixed at the density 0.12 fm$^{-3}$. It has been shown in \cite{camille11} that
phenomenological models fitted to the properties of nuclei and nuclear matter
have similar values of the symmetry energy for this density.  We generate a
set of models that differ in their value of the symmetry energy and
corresponding slope at saturation as indicated in Table~\ref{tab-sets-grho},
but have the same isoscalar properties. In Fig.~\ref{ener-sym}b) we show the
symmetry energy density dependence of this set of models. The set 1 is the
parametrization  IUFSU. The other parametrizations have a larger symmetry
energy and a larger slope $L$ at saturation. The maximum value of $L$ we have
considered is within the experimental values obtained from isospin diffusion
in heavy ion reactions \cite{chen05}.  The range of values considered for
$L$ span all the interval obtained for $L$ from different analysis of
experimental measurements \cite{slope} and a microscopic Brueckner
Hartree-Fock calculation \cite{isaac09}.


\begin{figure*}[ht]
  \centering
\begin{tabular}{cc}
\includegraphics[width=8cm,height=6.0cm,angle=0]{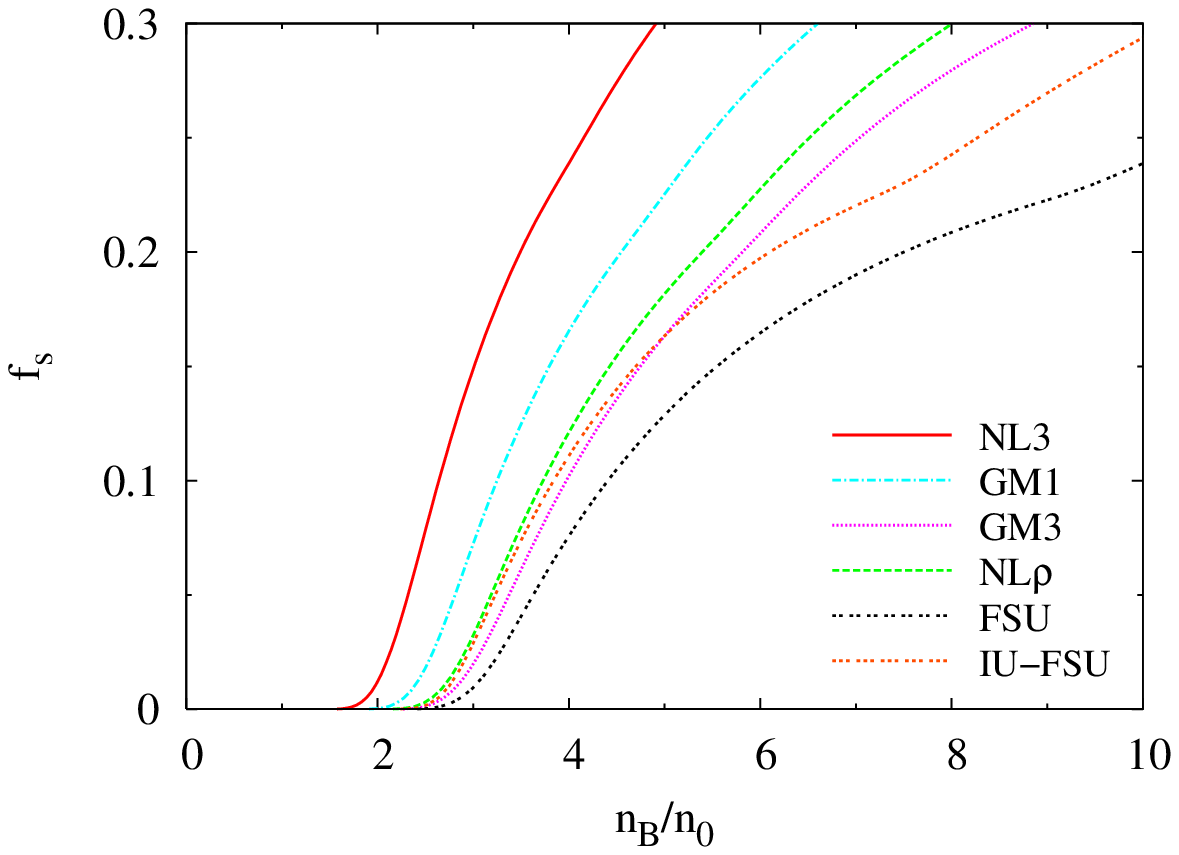} 
 &
\includegraphics[width=8cm,height=6.0cm,angle=0]{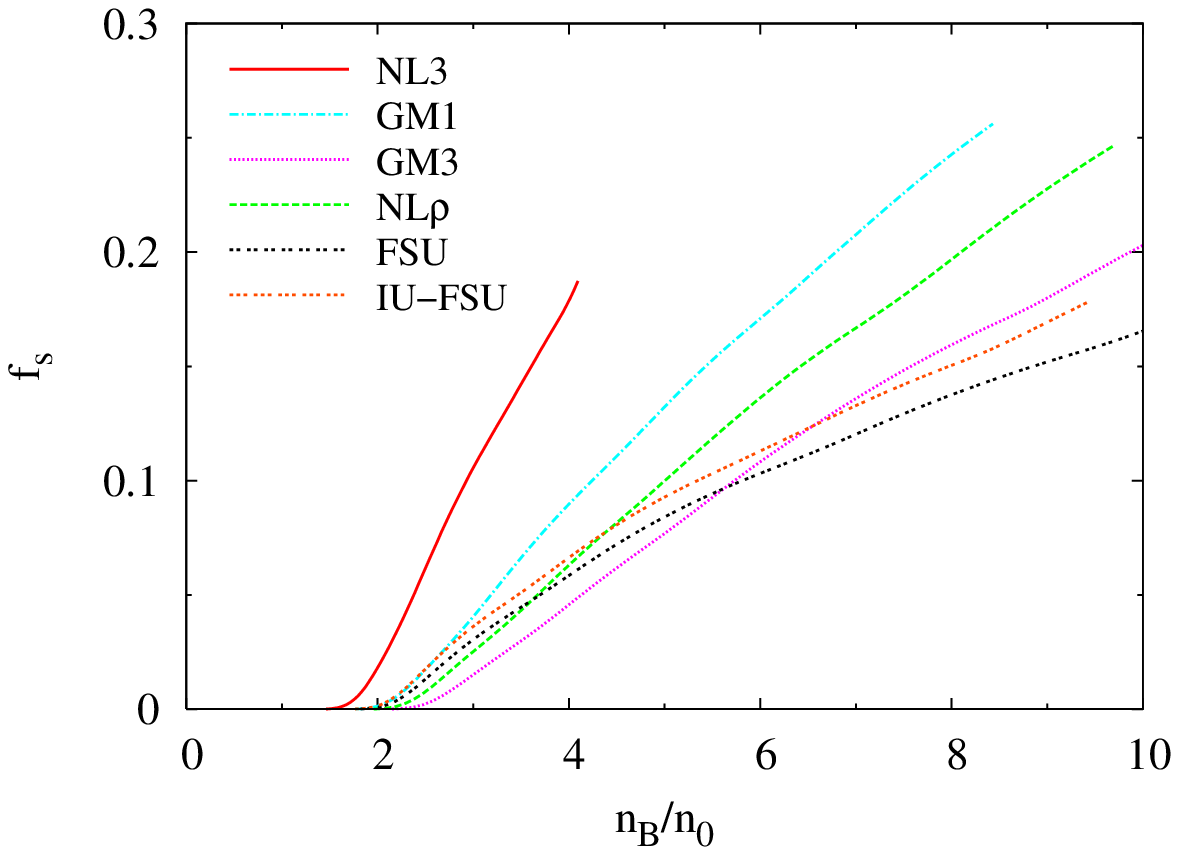} 
\\ {\bf (a)} & {\bf (b)}
\end{tabular}
\caption{(Color online) Strangeness fraction when hyperons are present: (a) Set A, (b) Set B. }
 \label{fig-hyp34}
\end{figure*}

\begin{figure*}[ht]
  \centering
\begin{tabular}{cc}
\includegraphics[width=8cm,height=6.0cm,angle=0]{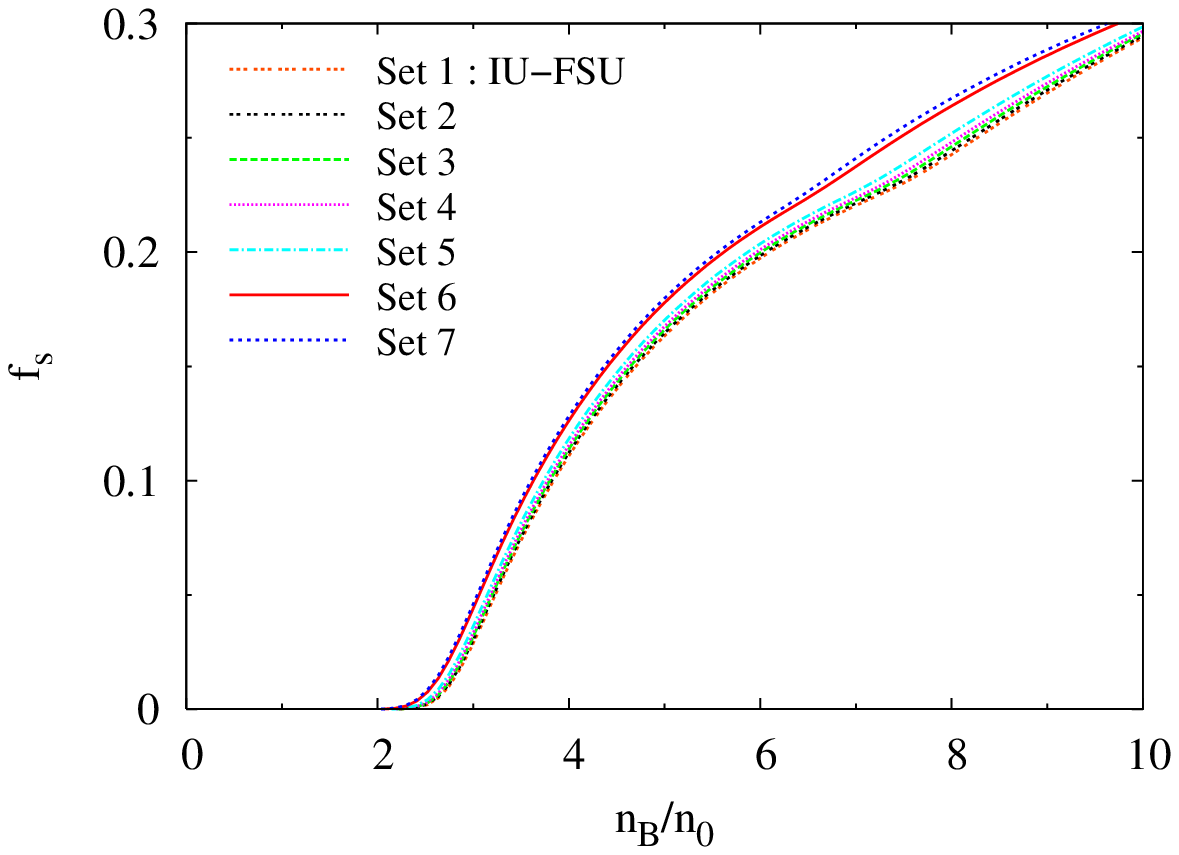} &   
\includegraphics[width=8cm,height=6.0cm,angle=0]{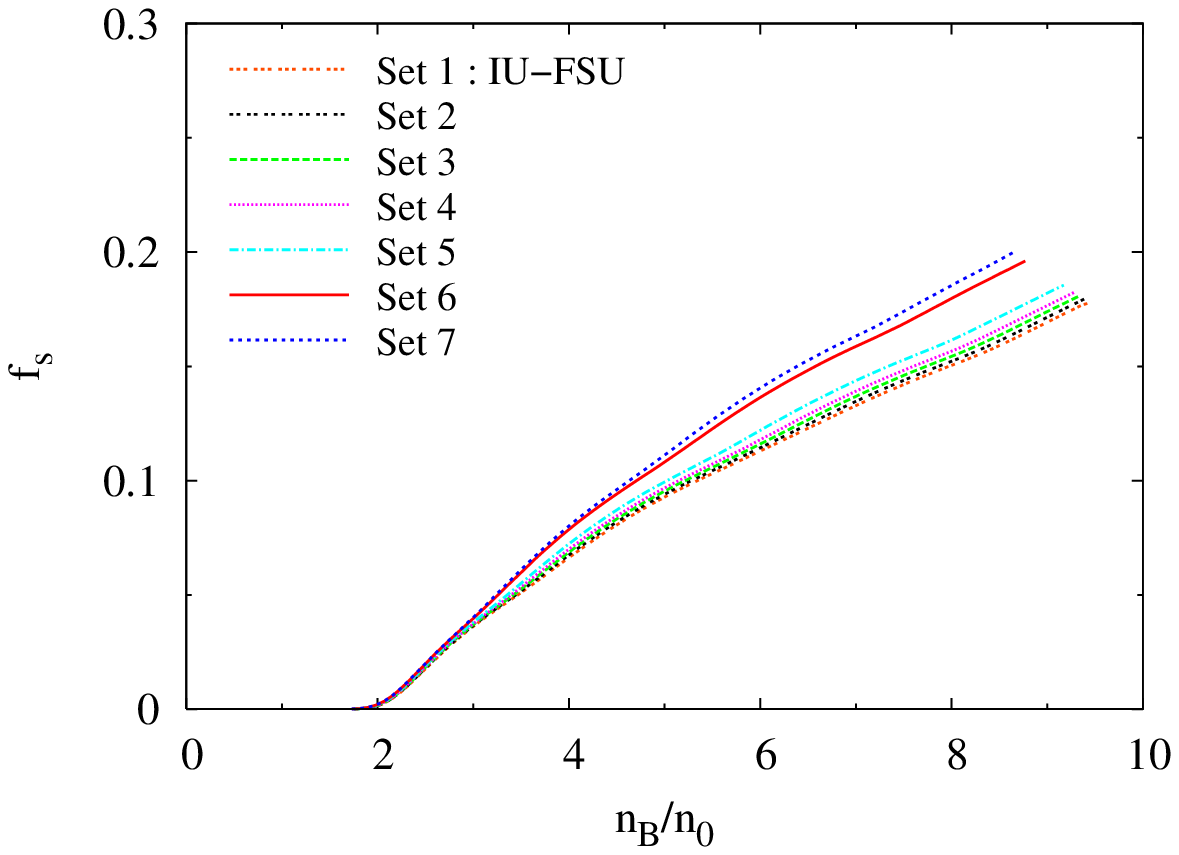}    
\\ {\bf (a)} & {\bf (b)}
\end{tabular}
\caption{(Color online) Strangeness fraction for IUFSU and related parametrizations with 
a) meson-hyperon coupling set A; b) meson-hyperon coupling set B; }
 \label{fig-hyp-sym12}
\end{figure*}

\myindent The effect of the symmetry energy on the strangeness fraction is seen in
 Figs. \ref{fig-hyp34} and \ref{fig-hyp-sym12}. 
It is clear from these figures that the strangeness content is
  sensitive to the model and the meson-hyperon couplings. The softer the EOS
  the larger the strangeness onset density and the smaller the strangeness
  content. A large meson-hyperon vector coupling, as occurs in set B, hinders
  the formation of hyperons.

\begin{figure*}[ht]
  \centering
\begin{tabular}{cc}
\includegraphics[width=0.75\linewidth,angle=0]{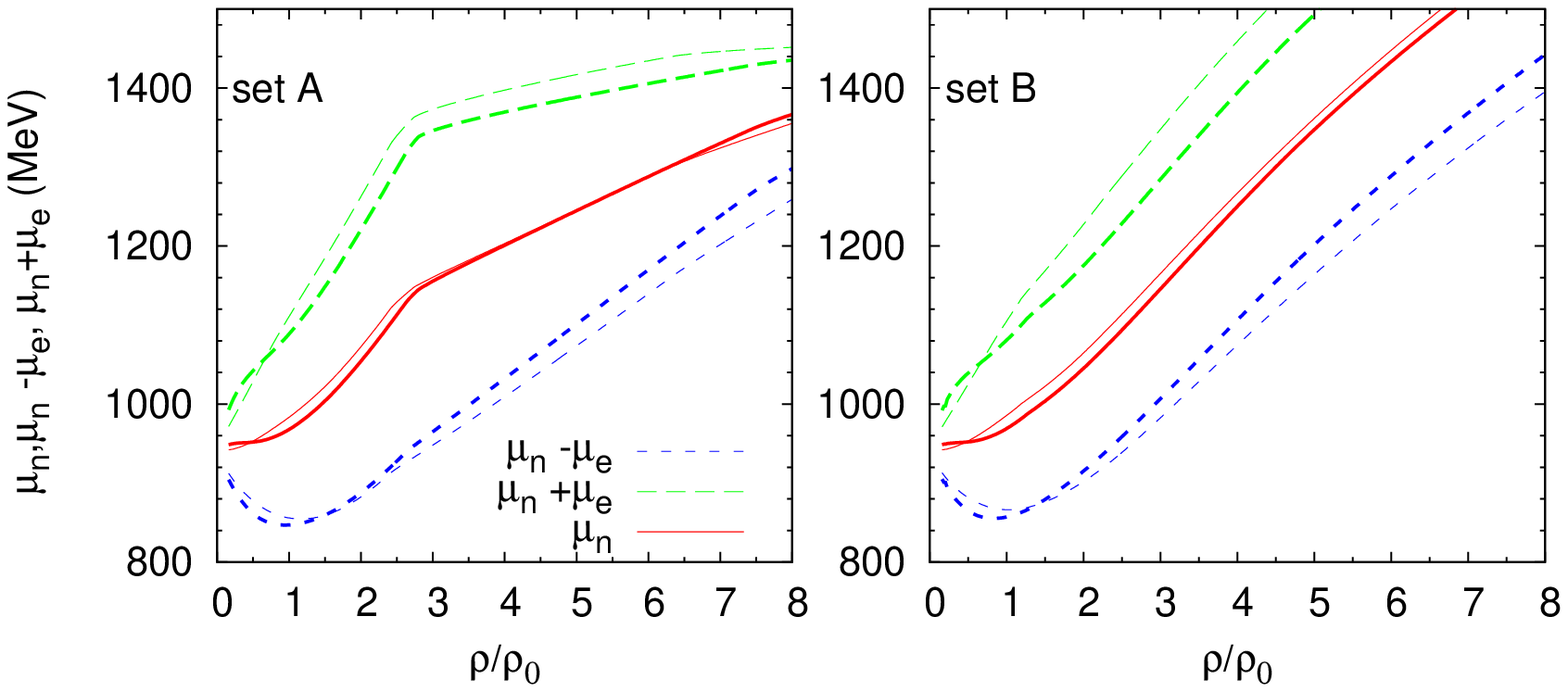}   
\end{tabular}
\caption{(Color online) Chemical potential for neutral (full lines), positively
  charged (dotted lines) and
  negatively charged (dashed lines) baryons in $\beta$ equilibrium: hyperon set A (left
  panel) and  hyperon set B (right panel). The thick lines are for IU-FSU (set
  1) and the thin lines for set 7 of the modified IU-FSU.}
 \label{chem}
\end{figure*}

The symmetry energy is
 affecting directly the isovector chemical potential and, therefore, the
 chemical equilibrium. In Fig. \ref{chem} the  chemical potential for neutral, positively
  charged  and
  negatively charged baryons in $\beta$ equilibrium are represented. It is
  seen that the neutron chemical potential becomes slightly smaller for a
  softer symmetry energy, because the $\rho$-meson field is 
weaker. However, the density dependence of the symmetry energy has a
  stronger effect on the electron chemical potential: a smaller $L$
  corresponds to a smaller proton fraction and, therefore, electron fraction
  so that the electron chemical potential decreases when $L$ decreases. As
  a result the sum $\mu_n+\mu_e$, which defines the chemical potential of
  single negatively charged baryon, feels a much stronger reduction than the
  neutron chemical potential, and the difference  $\mu_n-\mu_e$, which defines the chemical potential of
  single positively charged baryons, increases above saturation
  density when $L$ decreases.

\begin{figure*}[ht]
  \centering
\begin{tabular}{cc}
\includegraphics[width=8cm,height=6.0cm,angle=0]{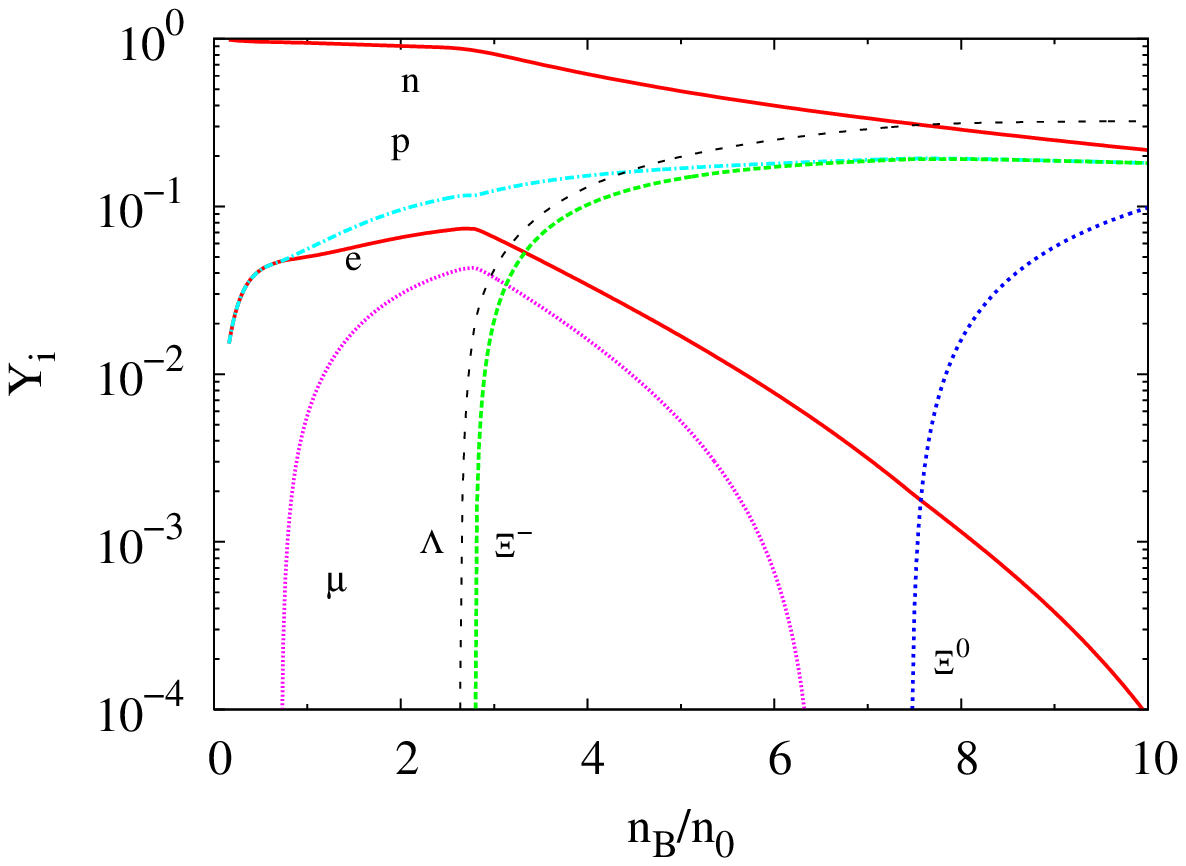}  
 &
\includegraphics[width=8cm,height=6.0cm,angle=0]{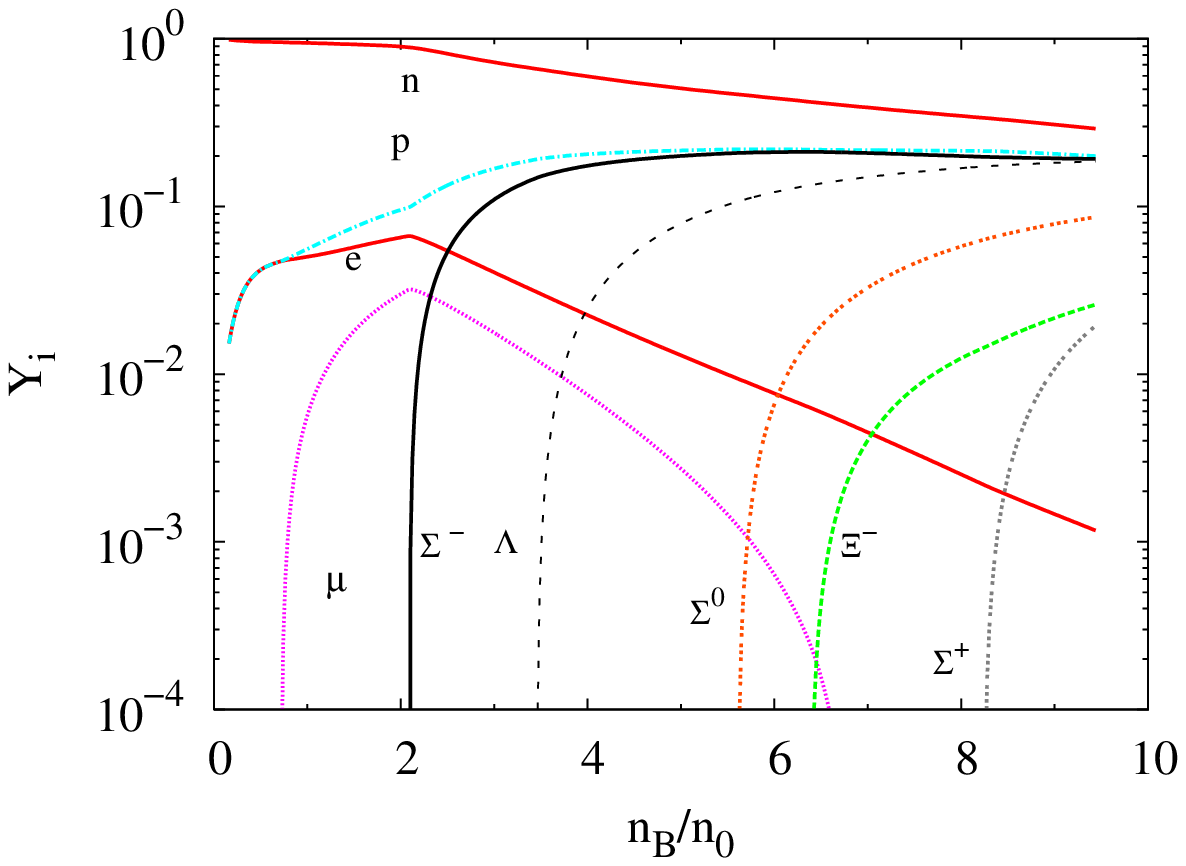}  
\\ {\bf (a)} & {\bf (b)}
\end{tabular}
\caption{(Color online) Particle fractions, IU-FSU a) Set A; b) Set B }
 \label{fig-hyp-fraction}
\end{figure*}

\myindent As a consequence, a soft symmetry energy  shifts the hyperon onset to
 larger densities, if $\Lambda$, a neutral hyperon with isospin zero, is the first hyperon to
 appear. This is the case of set A (see the left panel of Fig. \ref{fig-hyp-fraction}). 
 However,  the onset of hyperons is not affected by
 $L$ when a negatively charged hyperon such as $\Sigma^-$ is the first hyperon
 to appear as for instance in set B  (see the right panel of Fig. \ref{fig-hyp-fraction}). Although the sum $\mu_n+\mu_e$ decreases, the same happens with the $\Sigma^-$ chemical potential and the net
result is an almost independence of the onset of the hyperon $\Sigma^-$ on
$L$. As
 soon as the $\Lambda$ also sets on the different parametrizations of the modified
 IU-FSU start to differ.

\myindent In Tables \ref{tab-tov-nohyp}-\ref{tab-tovC} the direct Urca onset density,
mass and radius of maximum mass
stars, the radius of stars with $M=1.0 \, M_\odot$ and 1.4 $M_\odot$, the
symmetry energy ${\cal E}_{sym}$ and the symmetry energy slope $L$ are given for
a) stars without hyperons, Tables  \ref{tab-tov-nohyp} and \ref{tab-tov-nohyp-esym},  
b) stars with hyperons that couple to mesons through set A,
Tables \ref{tab-tovA.varios} and \ref{tab-tovA}, and  through set B,
Tables \ref{tab-tovC.varios} and \ref{tab-tovC}. The properties of stars
obtained with models NL3, GM1, GM3,
NL$\rho$, FSU and IUFSU are given in Tables  \ref{tab-tov-nohyp}, \ref{tab-tovA.varios}
and \ref{tab-tovC.varios}, and the properties of stars from a set of EOS obtained from  IUFSU by changing
the isovector channel are presented in Tables \ref{tab-tov-nohyp-esym}, \ref{tab-tovA}
and \ref{tab-tovC}.

\myindent Cooling of the star by neutrino emission can occur relatively fast if the
direct Urca process, $n\to p+e^-+ \bar{\nu}_e$, is allowed \cite{urca}.
The direct Urca (DU) process takes place when the proton fraction exceeds a critical 
value $x_{\rm \scriptscriptstyle DU}$\cite{urca}, which can be evaluated in terms of the leptonic
fraction as \cite{klaen06}:
\begin{equation}
x_{\rm \scriptscriptstyle DU} = \frac{1}{1 + (1 + x_e^{1/3})^3}  \:.
\end{equation}
$ $

where $x_e = n_e/(n_e + n_{\mu})$ is the leptonic fraction, $n_e$ is the number 
density of electrons and $n_{\mu}$ is the number density of muons. Cooling
rates of neutron stars seem to indicate that this fast cooling process does
not occur and, therefore, a constraint is set imposing that the direct Urca
process is only allowed in stars with a mass larger than $1.5 \, M_\odot$, or a
less restrictive limit,  $1.35 \, M_\odot$ \cite{klaen06}. Since the onset of
the direct Urca process is closely related with the density dependence of the
symmetry energy, this constraint gives information on the isovector channel of
the EOS.

\begin{figure*}[ht]
  \centering
\begin{tabular}{cc}
\includegraphics[width=8cm,height=6.0cm,angle=0]{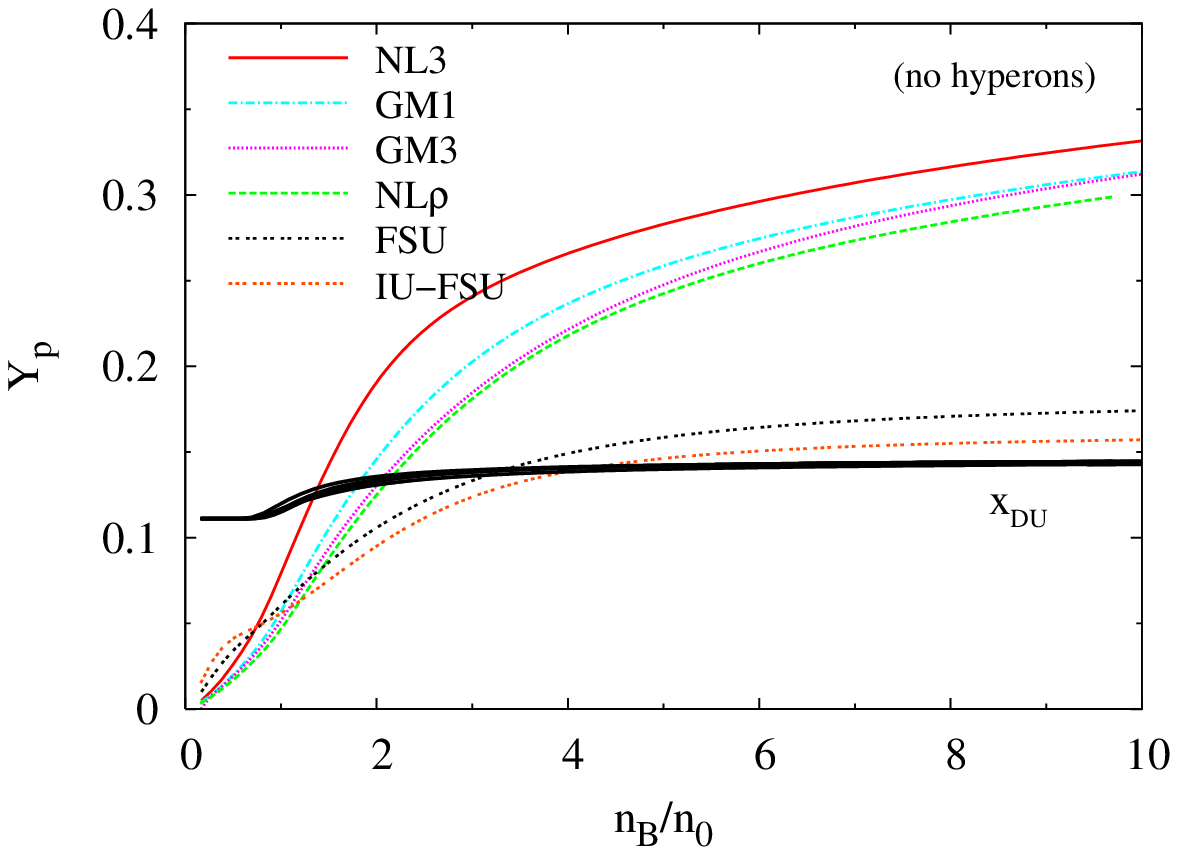}  
 &
\includegraphics[width=8cm,height=6.0cm,angle=0]{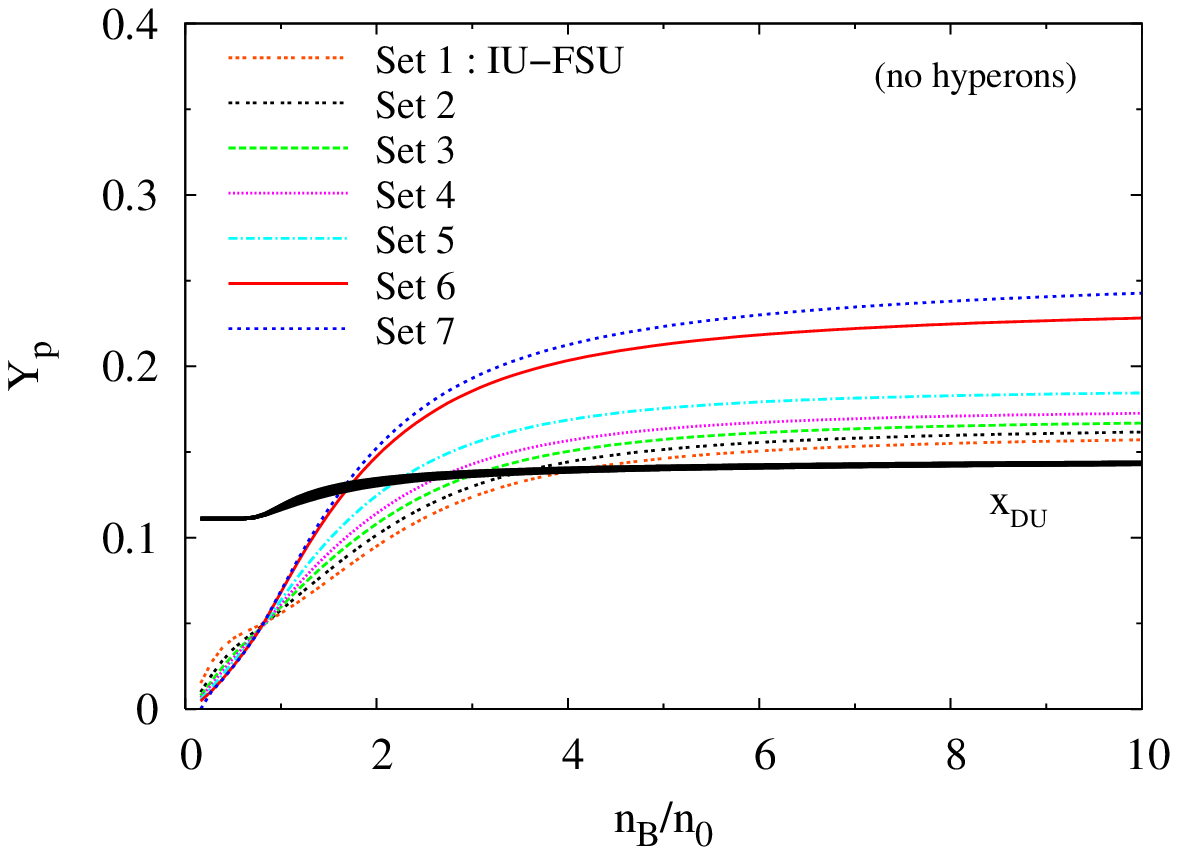}  
\\ {\bf (a)} & {\bf (b)}
\end{tabular}
\caption{(Color online) Onset of the direct Urca process in stellar matter without hyperons:
 proton fraction for $\beta$-equilibrium matter and proton fraction at the onset of
  the direct Urca process (black region) a) for NL3, GM1, GM3, NL$\rho$, FSU, IU-FSU, b) modified IU-FSU.
}
 \label{fig-nohyp12}
\end{figure*}

\myindent In Fig.\ref{fig-nohyp12} the proton
  fractions for $\beta$-equilibrium matter are plotted for NL3, GM1, GM3, FSU
  and IUFSU. The black region defines the proton fraction at the onset of
  the direct Urca process.

\begin{figure*}[ht]
  \centering
\begin{tabular}{cc}
\includegraphics[width=0.75\linewidth,angle=0]{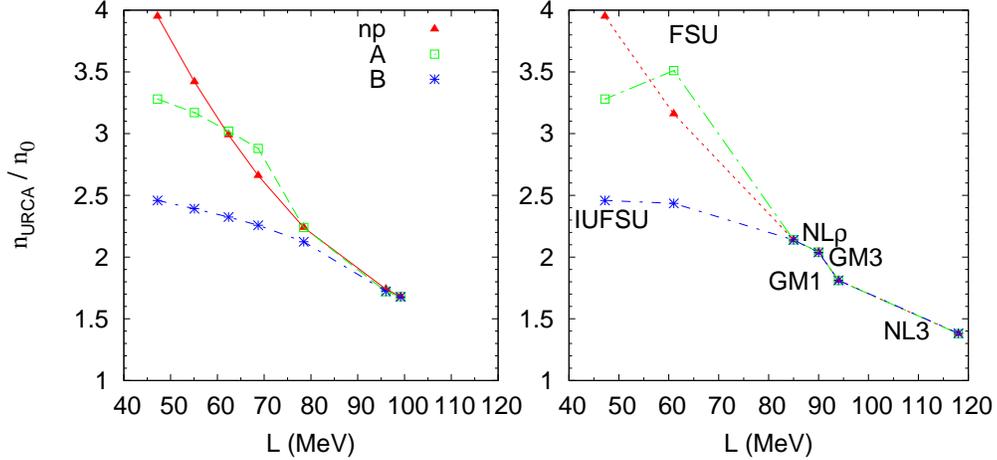}  
\end{tabular}
\caption{(Color online) Onset of the direct Urca process in stellar matter.
  Left panel  modified IUFSU model and right panel NL3, GM1, GM3, NL$\rho$,
  FSU and IUFSU, for nohyperon matter (red triangles), hyperon
coupling set A (green  squares) and  hyperon
coupling set B (blue stars).
}
 \label{urca}
\end{figure*}

\myindent The effect of the symmetry energy and the hyperon content on the onset density
of the direct Urca process is seen in figures \ref{urca} as function of the slope $L$  for the IUFSU and
modified versions in the left panel , and for   NL3, GM1, GM3,
NL$\rho$, FSU and IUFSU models in the right panel. We first analyse the
effect of the symmetry energy slope on this quantity. We conclude that:
a) for matter without hyperons the larger the $L$ the smaller the
neutron-proton asymmetry above the saturation density and, therefore, the
smaller the direct Urca onset density; 
a) the larger the slope the smaller the onset density because a larger $L$
corresponds to a harder symmetry energy and, therefore, larger fractions of
protons are favored;  
c) for a low value of $L$ the presence of
hyperons decreases the onset density. The effect of the inclusion of hyperons
depends on the hyperon-meson coupling. With Set A,  $\Lambda$ is the first hyperon
to appear as can be seen in Fig.~\ref{fig-hyp-fraction} a). With the onset of the $\Lambda$, the neutron
fraction decreases as well as the proton fraction. The behavior of the onset
density for the DU depends on the balance between these two effects.
 In general  the DU process is favored but for a small range of $L$
 ($65<L<80$ MeV), the DU process may occur at densities larger than the values obtained for
nucleonic matter. 

\myindent For set B,  $\Sigma^-$ is the first hyperon to appear according to 
Fig.~\ref{fig-hyp-fraction} b). 
With the onset of a
negatively charged hyperon, $\Sigma^-$ or $\Xi^-$, there is an increase of the
proton fraction due to electrical neutrality as well as a decrease of the
neutron fraction: both effects favor the
DU onset.

\begin{figure*}[ht]
  \centering
\begin{tabular}{cc}
\includegraphics[width=8cm,height=6.0cm,angle=0]{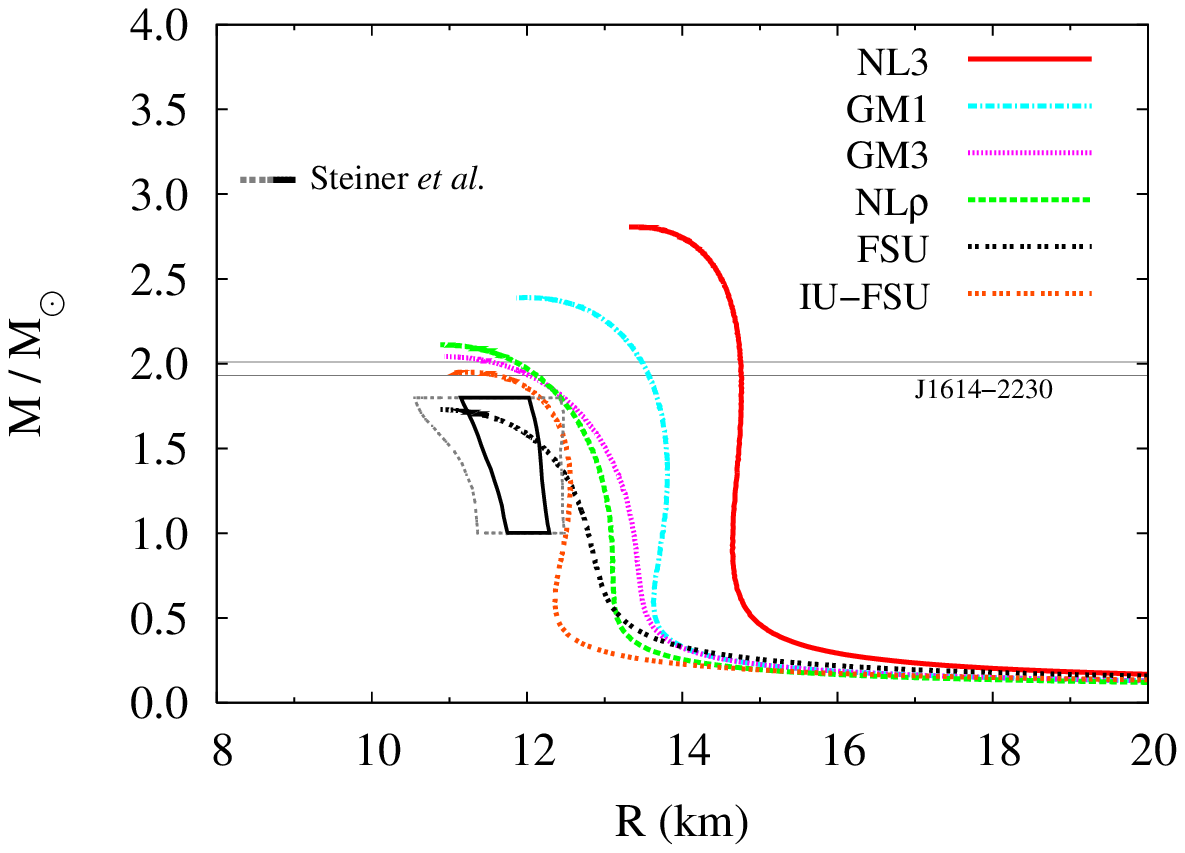}  
 &
\includegraphics[width=8cm,height=6.0cm,angle=0]{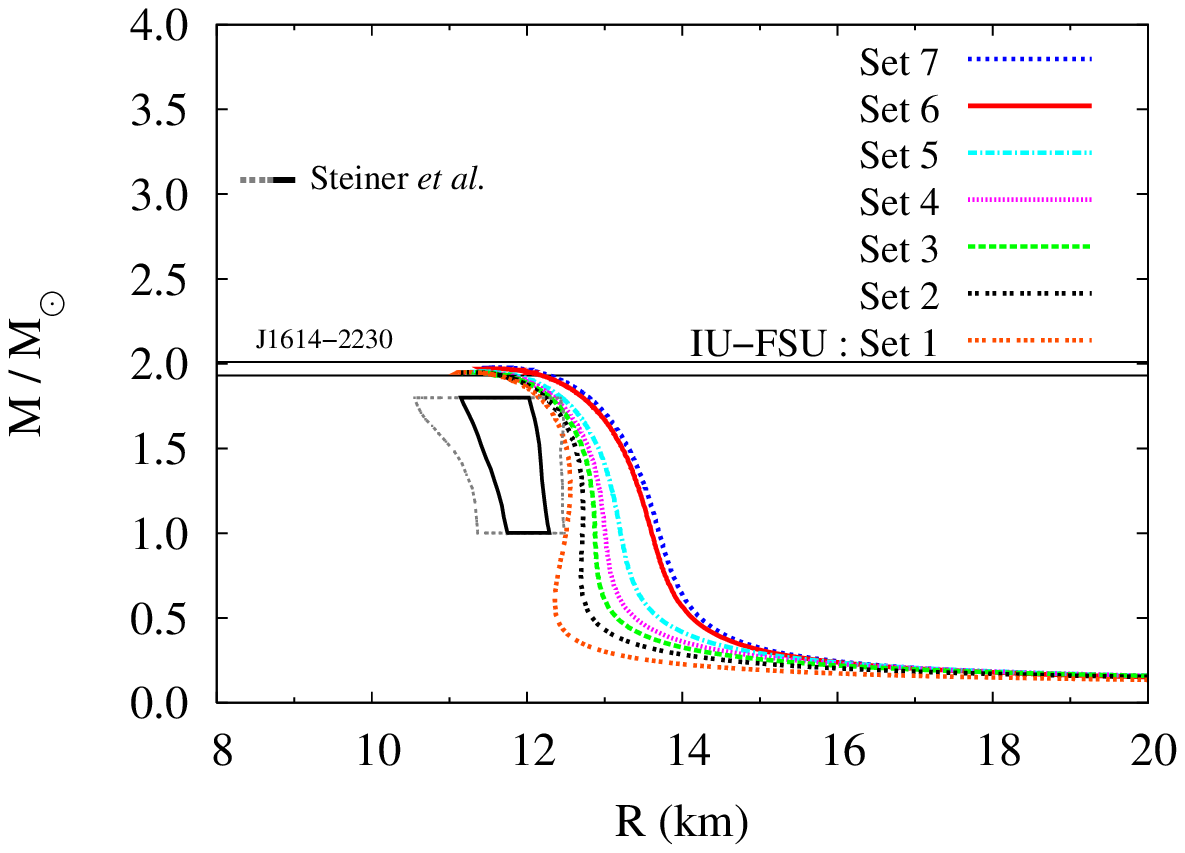}  
\\ {\bf (a)} & {\bf (b)}
\end{tabular} 
\caption{(Color online) Mass-radius curves obtained for no-hyperon stellar
  matter EOS: a) NL3, GM1, GM3, NL$\rho$, FSU and IUFSU b) IU-FSU and  modified IU-FSU. The areas limited by 
the dotted grey and solid black curves are given by \cite{steiner10}.}
 \label{fig-nohyp23}
\end{figure*}

\begin{figure*}[ht]
  \centering
\begin{tabular}{cc}
\includegraphics[width=8cm,height=6.0cm,angle=0]{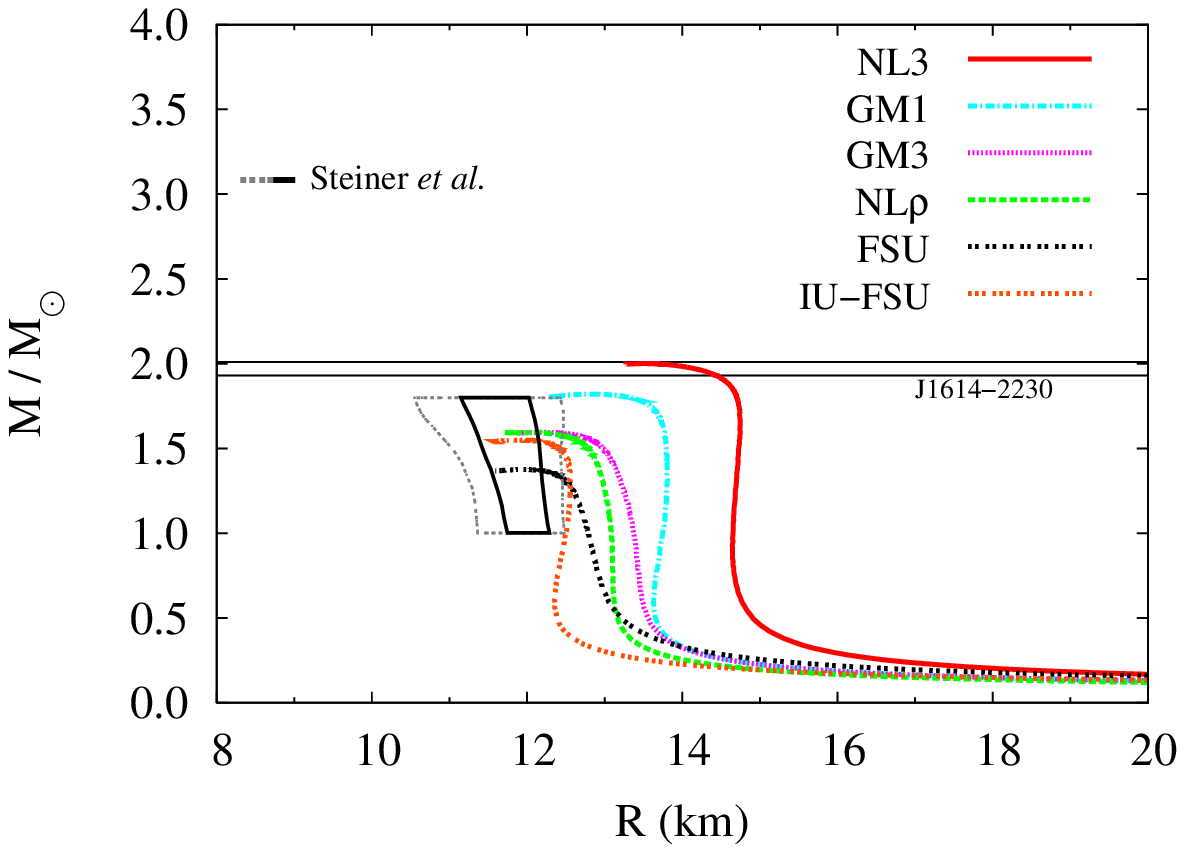}  
 &
\includegraphics[width=8cm,height=6.0cm,angle=0]{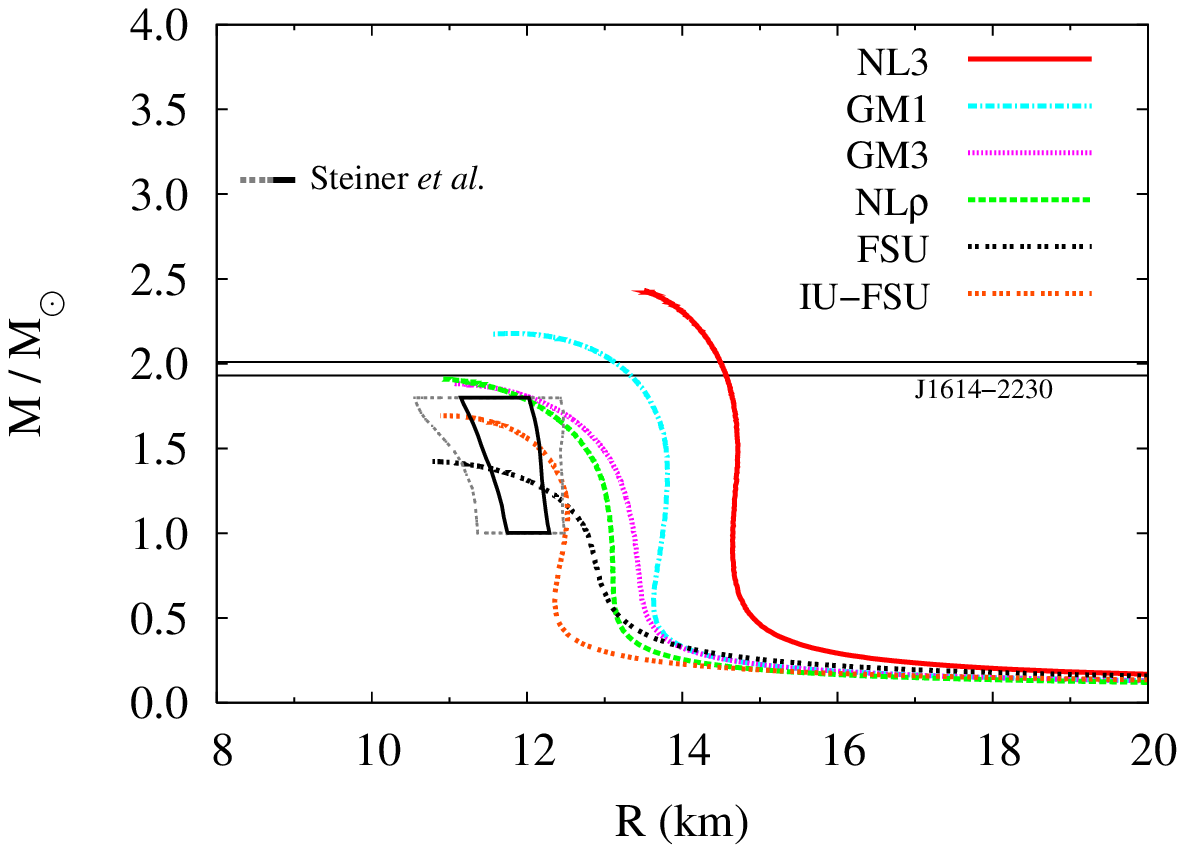}  
\\ {\bf (a)} & {\bf (b)}
\end{tabular}
\caption{(Color online) Mass-radius relations obtained for NL3, GM1, GM3,
  NL$\rho$, FSU and IUFSU  with the hyperon-meson coupling (a)
  set  A; , (b) set  B. }
 \label{fig-hyp56}
\end{figure*}

\begin{figure*}[ht]
  \centering
\begin{tabular}{cc}
\includegraphics[width=8cm,height=6.0cm,angle=0]{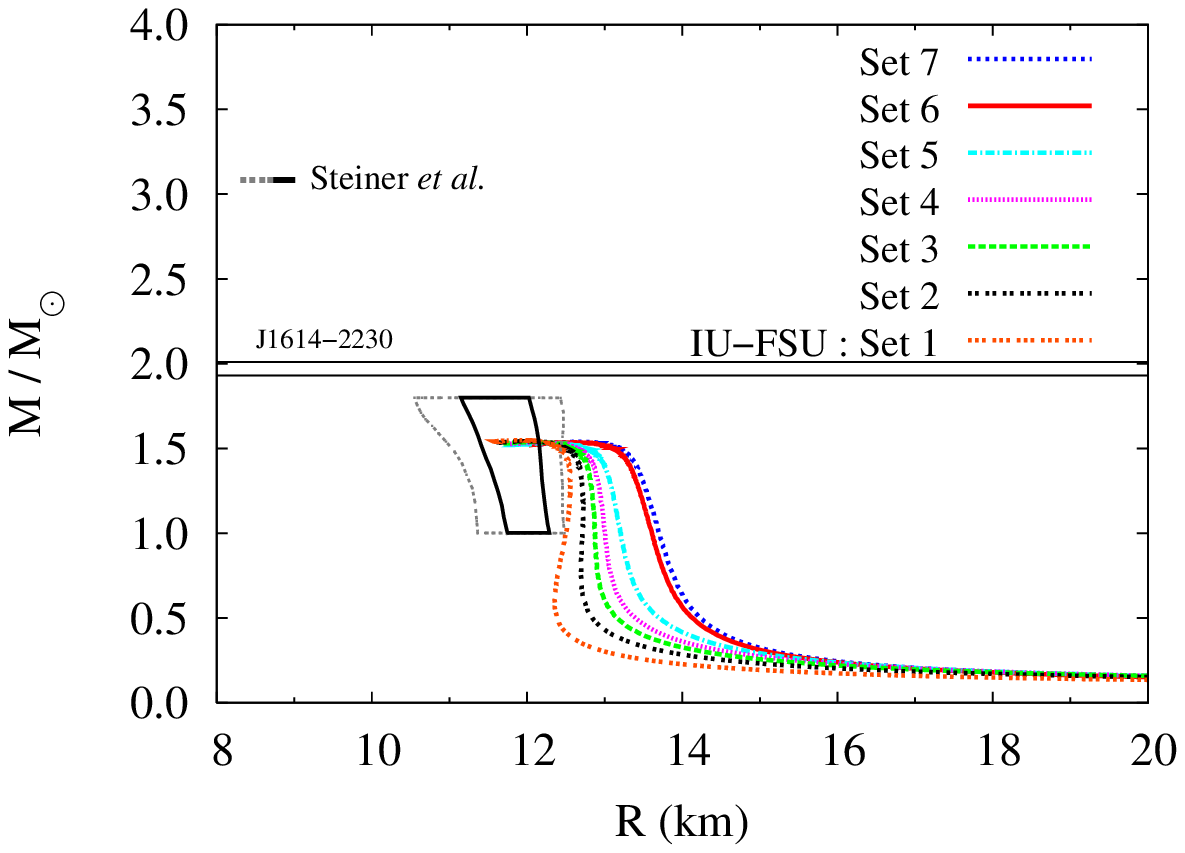}  
 &
\includegraphics[width=8cm,height=6.0cm,angle=0]{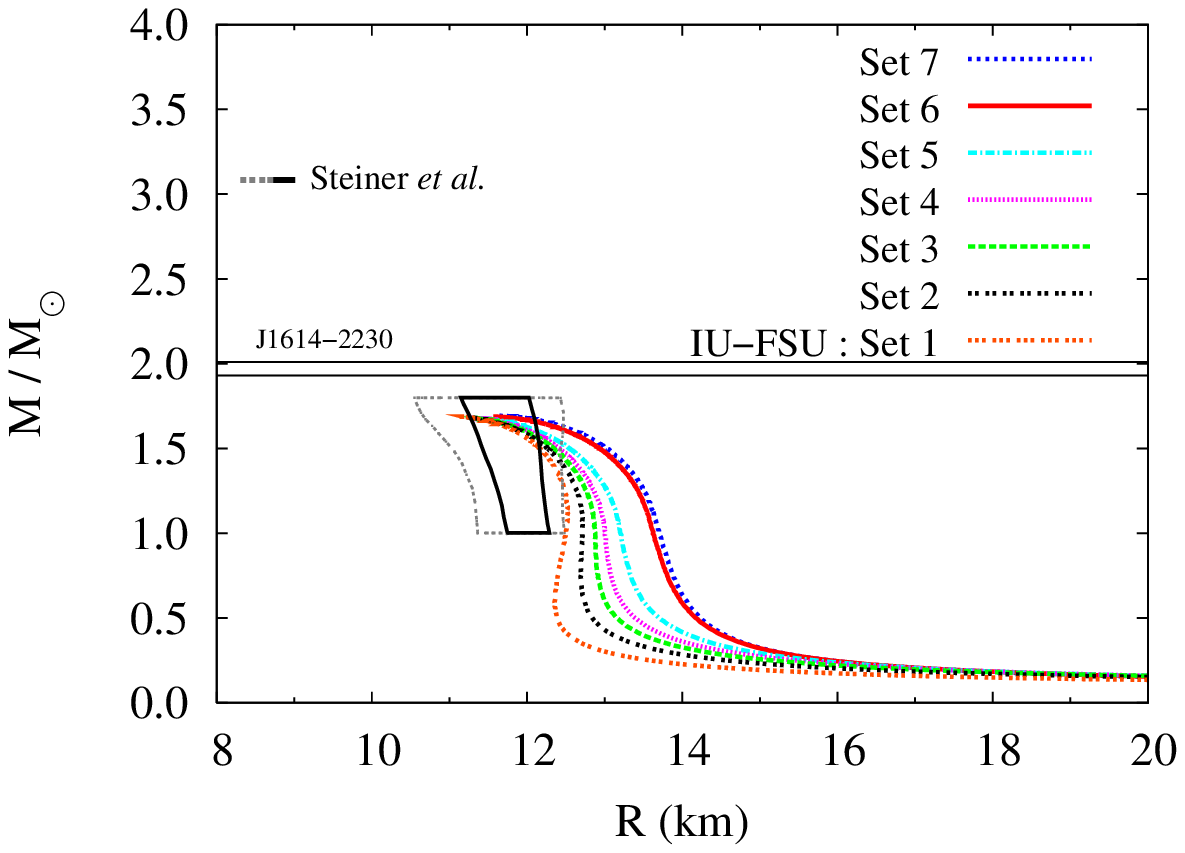}  
\\ {\bf (a)} & {\bf (b)}
\end{tabular}
\caption{(Color online) Mass-radius relations obtained with the modified IUFSU with the
  meson-hyperon coupling a) Set A; b) Set B }
 \label{fig-hyp-sym34}
\end{figure*}

\myindent In figures \ref{fig-nohyp23},  \ref{fig-hyp56} and \ref{fig-hyp-sym34} the
mass-radius for the families of stars obtained, respectively, from EOS without
hyperons and from  EOS with the meson-hyperon sets A and B are shown. In the
three figures we show on the left panel the curves obtained with models NL3,
GM1, GM3, NL$\rho$, FSU and IUFSU, while in the right panel the curves for
IUFSU and the modified IUFSU models. We also include the constrains obtained
by \cite{steiner10} and the mass of the millisecond binary pulsar J1614-2230
\cite{demorest10}.


\begin{figure*}[ht]
  \centering
\begin{tabular}{ccc}
& \includegraphics[width=0.95\linewidth,angle=0]{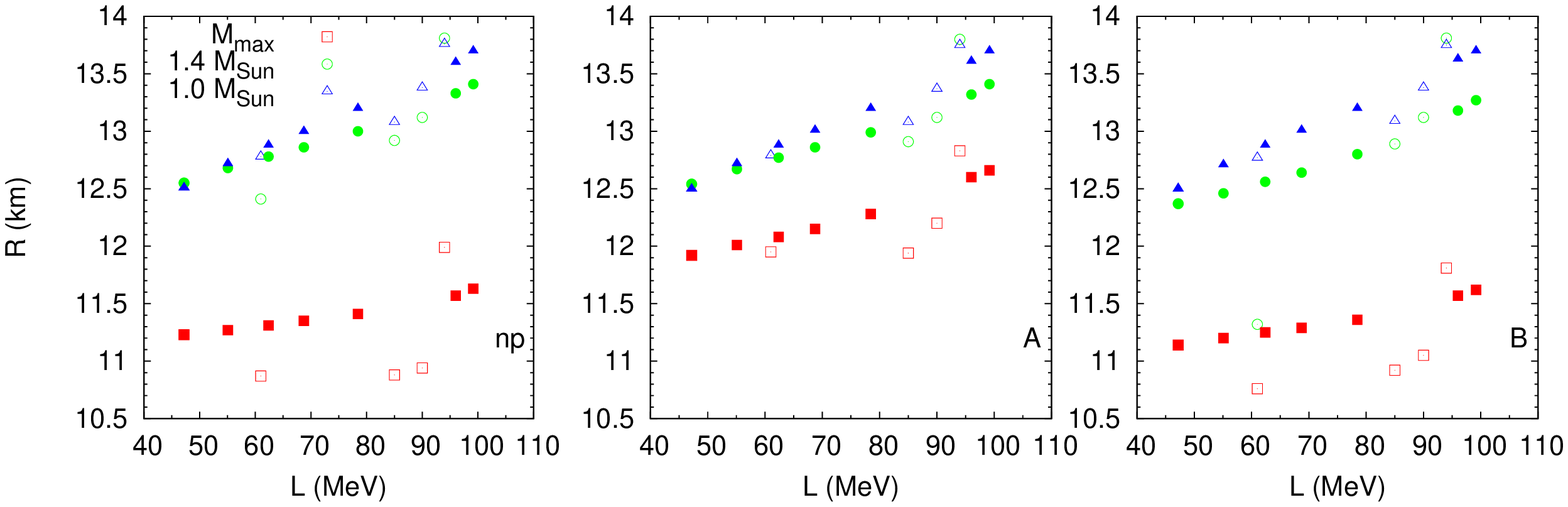} &  
\\  & \hspace*{0.45cm} {\bf (a)} \hspace*{4.5cm} {\bf (b)} \hspace*{4.5cm} {\bf (c)} & 
\end{tabular}
\caption{(Color online) Star radius dependence on the slope of the symmetry energy for
stars with maximum mass (squares), 1.4 $M_\odot$ (circles), 1  $M_\odot$ (triangles). The empty symbols are for the set of
different models considered with hyperon coupling A. The full symbols are for the modified IUFSU
a) no-hyperons ; b) with hyperons, coupling A; c) with hyperons, coupling B.
}
 \label{lradius}
\end{figure*}

\myindent We discuss first the results without hyperons, Fig. \ref{fig-nohyp23}. All the
models except FSU are able to describe the  pulsar J1614-2230. However, only
IUFSU and FSU satisfy the constrains on \cite{steiner10}. The set of relativistic 
mean field (RMF) models
chosen has quite different properties and is reflected in the differences
between the models: the harder models like NL3 and GM1 predict
larger masses and radius, the softer EOS, FSU, the smallest mass, the smaller
the $L$ the smaller the radius. This last property is clearly seen in the left
panel of Fig. \ref{lradius} were the radius of maximum mass stars (squares),
1.4 $M_\odot$ stars (circles) and 1.0$M_\odot$ stars (triangles) are plotted
as a function of the symmetry energy slope for nucleonic stars. The full
symbols are for the modified IUFSU models and the empty ones for the NL3, GM1,
GM3, NL$\rho$, FSU models. The modified IUFSU models show that if the
isoscalar channel is left unchanged the radius decreases if $L$
decreases. This reduction is larger for 1.0$M_\odot$ stars (more than 1 km for
$45<L<100$ MeV) but even for the maximum mass configurations there is still a
0.5 km difference. The set of RMF models chosen also show the same
trend. However, since the isoscalar properties differ among the
models, and they also affect the radius, the linear behavior is not present. 

\myindent  Including hyperons in the EOS makes the EOS softer at large densities and the
 mass of the maximum mass stars is smaller \cite{prakash97}. This is seen in Fig. \ref{fig-hyp56}
were the mass-radius curves obtained with hyperon-meson coupling set A and B
are plotted, respectively, in the left and right panels. We conclude that
it is important to have correct couplings since the star masses are sensitive
to the hyperon couplings. For set A only the NL3 model is able to
describe the PSR J1316-2230, while within set B NL3, GM1, GM3 and NL$\rho$ are
able to describe a star with a mass 1.97 $\pm$ 0.4 $M_\odot$. The trend
discussed above
between the star radius and $L$ is still present in these stars, see the empty
symbols in the middle and right panels of Fig. \ref{lradius}. This trend is
confirmed  by the modified IUFSU models,  full
symbols in the middle and right panels of Fig. \ref{lradius}. Stars with 1.0
and 1.4 $M_\odot$ contain no hyperons, or only a small fraction and therefore
their radius do not different from the results obtained for $np$ matter.
Maximum mass stars, however, do have hyperons and their radius depend on the
hyperon couplings chosen: for set A radius is larger and the maximum mass is
smaller than the corresponding quantities predicted by set B. None of the
models is able to describe the PSR J1614-2230. We also conclude that the mass of the
maximum mass configuration is quite insensitive to the symmetry energy slope. 



\section{Conclusions}

In the present work we have studied the effect of the density dependence of
the symmetry energy 
on the star properties, namely the  hyperon content, DU,
raius and mass.

\myindent The study was performed within the RMF framework. We have considered parametrizations which have
been fitted to the equilibrium properties of stable and unstable nuclei  and/or
dynamical response of nuclei, or to the saturation properties of symmetric
nuclear matter. In order to test the density dependence of the symmetry energy
we have considered  the IU-FSU parametrization. Keeping the isoscalar
channel fixed, we changed the isovector channel in order to reproduce the
values of the symmetry energy slope which have been obtained from experimental
measurements, $40< L< 110$ MeV \cite{chen05,slope}.
 For the hadronic EOS we have considered two
different parametrizations of the meson-hyperon couplings: in set A we
considered the couplings that reproduce the binding energy of hyperons to
symmetric nuclear matter; however, since only the binding energy of the
$\Lambda$ is well determined, we have also considered set B, which corresponds
to the couplings proposed in \cite{gm1} with  $x_s=0.8$. With set B and the
GM1 parametrization of the RMF for the nuclear EOS, the
authors of \cite{bombaci08} could obtain a maximum mass star configuration
of the order of the one measured by Demorest {\it et al.} \cite{demorest10}
recently.

\myindent 
We have analyzed the effect of $L$ on the radius of stars with 1 $M_\odot$,
1.4 $M_\odot$ and maximum mass configurations. The first two cases correspond
to stars that have a central baryonic density in the range $1.5\rho_0 - 3
\rho_0$ and therefore give information on the EOS just above the saturation
densities. These densities will be probed at FAIR \cite{fair-gsi}.
We have concluded that the radius of the star is sensitive to the slope $L$,
and, in particular, the smaller the value of the slope, the smaller the radius of the
star.
It was also shown that the density dependence of the symmetry energy affects
the onset density of the direct Urca process: the smaller $L$, the larger the
density. This can be understood because a smaller $L$ at saturation
corresponds to a softer symmetry energy at high densities, and, therefore, a
smaller proton fraction.  However, the DU onset also depends on the
hyperon content and the hyperon-meson couplings. If $\Lambda$ 
is the first hyperon to appear, the DU may be hindered or favored
according to a balance between the  neutron and proton reductions. However, if 
a negatively charged hyperon such as the $\Sigma^-$ is the first hyperon to
appear, there is a decrease of the neutron fraction and an increase of
the proton fraction: both effects favor the DU onset.   

\myindent It was also shown that the larger $L$, the larger the hyperon content because 
a larger $L$ makes the EOS harder and therefore, it is energetically favorable to have a larger hyperon
fraction. However, this only occurs if the first hyperon to appear is the
$\Lambda$. A delicate balance between the increase suffered by the chemical
potential of the $\Sigma^-$ and the neutron plus electron chemical potential
may make the hyperon onset independent of $L$.
We also conclude that the total strangeness content is
sensitive to the meson-hyperon couplings and stronger constraints on the
determination of these constants are required. For larger hyperon-vector meson
couplings we have obtained a smaller strangeness content.
According to recent estimates, based on a microscopic non-relativistic approach, including
hyperon degrees of freedom seems to make the EOS too soft even including
3-body forces \cite{vidana11}, so that at most
star masses of 1.6 $M_\odot$ are attained, far from the very precise mass
value recently measured \cite{demorest10}. More data on hypernuclei are required 
in order to clarify this point.


\begin{table*}[ht]
\centering
\begin{tabular}{ccccccccc}
\hline
\,\, Sets \,\, &\,\, $n_{\rm urca}/n_0 \,\,$ & \, $L$~(MeV) \,  &  \, ${\cal E}_{sym}$~(MeV) \,  &  \, $M_{\rm max}/M_{\odot}$ \,  &  \, $R_{M_{\rm max}}$~(km) \,  &  \, $R_{1.4M_{\odot}}$~(km) \,  &  \, $R_{1.0M_{\odot}}$~(km) \,  \\ 
\hline
NL3       &    1.38   &    118.0   &   37.4  &   2.81   &  13.38  &   14.71   &  14.65    \\
GM1       &    1.81   &     94.0   &   32.5  &   2.39   &  11.99  &   13.81   &  13.76    \\
GM3       &    2.04   &     90.0   &   32.5  &   2.04   &  10.94  &   13.12   &  13.38    \\
NL$\rho$  &    2.14   &     85.0   &   30.5  &   2.11   &  10.88  &   12.92   &  13.08    \\
FSU       &    3.16   &     61.0   &   32.6  &   1.73   &  10.87  &   12.41   &  12.78    \\
IU-FSU    &    3.95   &     47.2   &   31.3  &   1.95   &  11.23  &   12.55   &  12.51    \\
\hline
\hline 
\end{tabular}
\caption{Symmetry energy and no-hyperon star properties for NL3, GM1, GM3, FSU, IU-FSU EOS. 
The onset density of the direct Urca process,
symmetry energy slope and symmetry energy at saturation, mass and radius of
the maximum mass configuration, and radius of $1.4M_{\odot}$ and
$1.M_{\odot}$ stars are given. }
\label{tab-tov-nohyp}
\end{table*}

\begin{table*}[ht]
\centering
\begin{tabular}{ccccccccc}
\hline
\,\, Sets \,\, &\,\, $n_{\rm urca}/n_0 \,\,$ & \, $L$~(MeV) \,  &  \, ${\cal E}_{sym}$~(MeV) \,  &  \, $M_{\rm max}/M_{\odot}$ \,  &  \, $R_{M_{\rm max}}$~(km) \,  &  \, $R_{1.4M_{\odot}}$~(km) \,  &  \, $R_{1.0M_{\odot}}$~(km) \,  \\ 
\hline
1 & 3.95 & 47.20 & 31.34  & 1.95 & 11.23 & 12.55 & 12.51 \\ 
2 & 3.42 & 55.09 & 32.09  & 1.95 & 11.27 & 12.68 & 12.72 \\ 
3 & 2.99 & 62.38 & 32.74  & 1.95 & 11.31 & 12.78 & 12.88 \\ 
4 & 2.66 & 68.73 & 33.26  & 1.95 & 11.35 & 12.86 & 13.00 \\ 
5 & 2.24 & 78.45 & 34.00  & 1.95 & 11.41 & 13.00 & 13.20 \\ 
6 & 1.74 & 96.02 & 35.21  & 1.97 & 11.57 & 13.33 & 13.60 \\ 
7 & 1.68 & 99.17 & 35.41  & 1.98 & 11.63 & 13.41 & 13.70 \\
\hline 
\end{tabular}
\caption{Symmetry energy and no-hyperon star properties for IU-FSU modified EOS. 
The onset density of the direct Urca process,
symmetry energy slope and symmetry energy at saturation, mass and radius of
the maximum mass configuration, and radius of $1.4M_{\odot}$ and
$1.M_{\odot}$ stars are given.}
\label{tab-tov-nohyp-esym}
\end{table*}

\begin{table*}[ht]
\centering
\begin{tabular}{ccccccccc}
\hline
\,\, Sets \,\, &\,\, $n_{\rm urca}/n_0 \,\,$ & \, $L$~(MeV) \,  &  \, ${\cal E}_{sym}$~(MeV) \,  &  \, $M_{\rm max}/M_{\odot}$ \,  &  \, $R_{M_{\rm max}}$~(km) \,  &  \, $R_{1.4M_{\odot}}$~(km) \,  &  \, $R_{1.0M_{\odot}}$~(km) \,  \\ 
\hline
NL3       &    1.38   &    118.0   &   37.4  &   2.00   &  13.51  &   14.71   &  14.65   \\
GM1       &    1.81   &     94.0   &   32.5  &   1.82   &  12.83  &   13.80   &  13.75    \\
GM3       &    2.04   &     90.0   &   32.5  &   1.595  &  12.20  &   13.12   &  13.37    \\
NL$\rho$  &    2.14   &     85.0   &   30.5  &   1.594  &  11.94  &   12.91   &  13.08    \\
FSU       &    3.51   &     61.0   &   32.6  &   1.375  &  11.95  &     -     &  12.79    \\
IU-FSU    &    3.28   &     47.2   &   31.3  &   1.55   &  11.92  &   12.54   &  12.50    \\
\hline
\hline 
\end{tabular}
\caption{Symmetry energy and star properties for NL3, GM1, GM3, FSU, IU-FSU EOS and set A
for the meson-hyperon couplings. The onset density of the direct Urca process,
symmetry energy slope and symmetry energy at saturation, mass and radius of
the maximum mass configuration, and radius of $1.4M_{\odot}$ and
$1.M_{\odot}$ stars are given.  }
\label{tab-tovA.varios}
\end{table*}

\begin{table*}[ht]
\centering
\begin{tabular}{ccccccccc}
\hline
\,\, Sets \,\, &\,\, $n_{\rm urca}/n_0 \,\,$ & \, $L$~(MeV) \,  &  \, ${\cal E}_{sym}$~(MeV) \,  &  \, $M_{\rm max}/M_{\odot}$ \,  &  \, $R_{M_{\rm max}}$~(km) \,  &  \, $R_{1.4M_{\odot}}$~(km) \,  &  \, $R_{1.0M_{\odot}}$~(km) \,  \\ 
\hline
NL3       &    1.3827   &    118.0   &   37.4  &   2.43   &  13.51  &   14.71   &  14.65   \\
GM1       &    1.8121   &     94.0   &   32.5  &   2.18   &  11.81  &   13.81   &  13.75    \\
GM3       &    2.0370   &     90.0   &   32.5  &   1.88   &  11.05  &   13.12   &  13.38    \\
NL$\rho$  &    2.1389   &     85.0   &   30.5  &   1.91   &  10.92  &   12.89   &  13.09    \\
FSU       &    2.4346   &     61.0   &   32.6  &   1.42   &  10.76  &   11.32   &  12.77    \\
IU-FSU    &    2.4585   &     47.2   &   31.3  &   1.69   &  11.14  &   12.37   &  12.50    \\
\hline
\hline 
\end{tabular}
\caption{Symmetry energy and star properties for NL3, GM1, GM3, FSU, IU-FSU EOS and set B
for the meson-hyperon couplings. The onset density of the direct Urca process,
symmetry energy slope and symmetry energy at saturation, mass and radius of
the maximum mass configuration, and radius of $1.4M_{\odot}$ and
$1.M_{\odot}$ stars are given.  }
\label{tab-tovC.varios}
\end{table*}

\begin{table*}[ht]
\centering
\begin{tabular}{ccccccccc}
\hline
\,\, Sets \,\, &\,\, $n_{\rm urca}/n_0 \,\,$ & \, $L$~(MeV) \,  &  \, ${\cal E}_{sym}$~(MeV) \,  &  \, $M_{\rm max}/M_{\odot}$ \,  &  \, $R_{M_{\rm max}}$~(km) \,  &  \, $R_{1.4M_{\odot}}$~(km) \,  &  \, $R_{1.0M_{\odot}}$~(km) \,  \\ 
\hline
1 & 3.28 & 47.20 & 31.34  & 1.55 & 11.92 & 12.54 & 12.50 \\ 
2 & 3.17 & 55.09 & 32.09  & 1.54 & 12.01 & 12.67 & 12.72 \\ 
3 & 3.02 & 62.38 & 32.74  & 1.54 & 12.08 & 12.77 & 12.88 \\ 
4 & 2.88 & 68.73 & 33.26  & 1.53 & 12.15 & 12.86 & 13.01 \\ 
5 & 2.24 & 78.45 & 34.00  & 1.53 & 12.28 & 12.99 & 13.20 \\ 
6 & 1.72 & 96.02 & 35.21  & 1.54 & 12.60 & 13.32 & 13.61 \\ 
7 & 1.68 & 99.17 & 35.41  & 1.54 & 12.66 & 13.41 & 13.70 \\
\hline 
\end{tabular}
\caption{Symmetry energy and star properties for IU-FSU modified EOS and set A
for the meson-hyperon couplings. The onset density of the direct Urca process,
symmetry energy slope and symmetry energy at saturation, mass and radius of
the maximum mass configuration, and radius of $1.4M_{\odot}$ and
$1.M_{\odot}$ stars are given.  }
\label{tab-tovA}
\end{table*}

\begin{table*}[ht]
\centering
\begin{tabular}{ccccccccc}
\hline
\,\, Sets \,\, &\,\, $n_{\rm urca}/n_0 \,\,$ & \, $L$~(MeV) \,  &  \, ${\cal E}_{sym}$~(MeV) \,  &  \, $M_{\rm max}/M_{\odot}$ \,  &  \, $R_{M_{\rm max}}$~(km) \,  &  \, $R_{1.4M_{\odot}}$~(km) \,  &  \, $R_{1.0M_{\odot}}$~(km) \,  \\ 
\hline
     1  &  2.4585 & 47.20 & 31.34   &   1.69   &   11.14   &   12.37    &   12.50     \\            
     2  &  2.3916 & 55.09 & 32.09   &   1.69   &   11.20   &   12.46    &   12.71     \\  
     3  &  2.3246 & 62.38 & 32.74   &   1.68   &   11.25   &   12.56    &   12.88     \\         
     4  &  2.2577 & 68.73 & 33.26   &   1.68   &   11.29   &   12.64    &   13.01     \\            
     5  &  2.1239 & 78.45 & 34.00   &   1.68   &   11.36   &   12.80    &   13.20     \\            
     6  &  1.7225 & 96.02 & 35.21   &   1.69   &   11.57   &   13.18    &   13.63     \\         
     7  &  1.6779 & 99.17 & 35.41   &   1.69   &   11.62   &   13.27    &   13.70     \\          
\hline 
\end{tabular}
\caption{Symmetry energy and star properties for IU-FSU modified EOS and set B
for the meson-hyperon couplings. The onset density of the direct Urca process,
symmetry energy slope and symmetry energy at saturation, mass and radius of
the maximum mass configuration, and radius of $1.4M_{\odot}$ and
$1.M_{\odot}$ stars are given.  }
\label{tab-tovC}
\end{table*}

\section*{ACKNOWLEDGMENTS}

This work was partially supported by CNPq (Brazil) and
CAPES(Brazil)/FCT(Portugal) under project 232/09, by FCT (Portugal) under the grants
PTDC/FIS/113292/2009 and CERN/FP/109316/2009. 
 R. C. is grateful for the warm hospitality at Centro de F\'{\i}sica Computacional/FCTUC.


\end{document}